**Data-Driven Micromechanical Characterization and Mapping of Shale Rocks Using High Speed Nanoindentation**

Shaziya A. Banu, S.M.ASCE[1], Xukai Zhang, M.ASCE[2], Samah A. Mahgoub[3], Christopher C. Walker[4], George M. Pharr, Ph.D.[5], Arash Noshadravan, M.ASCE[6], and Sara Abedi, M.ASCE[7]

[1]Graduate Assistant Research, Zachry Department of Civil and Environmental Engineering, Texas A&M University, College Station, TX-77840, USA; email: shazia_ahmeds@tamu.edu

[2]Postdoctoral Associate, Department of Civil and Environmental Engineering, Massachusetts Institute of Technology, Cambridge, MA 02139, USA; email: xukai@mit.edu

[3]Graduate Assistant Research, Zachry Department of Civil and Environmental Engineering, Texas A&M University, College Station, TX-77840, USA; email: samah.mahgoub@tamu.edu

[4]Graduate Assistant Research, Department of Materials Science and Engineering, Texas A&M University, College Station, TX-77840, USA; email: cwalke60@tamu.edu

[5]Professor, Department of Materials Science and Engineering, Texas A&M University, College Station, TX-77840, USA; email: pharr@tamu.edu

[6]Associate Professor, Zachry Department of Civil and Environmental Engineering, Texas A&M University, College Station, TX-77840, USA; email: noshadravan@tamu.edu (Corresponding author)

[7]Associate Professor, Harold Vance Department of Petroleum Engineering and affiliated with Zachry Department of Civil and Environmental Engineering, Texas A&M University, College Station, TX-77843, USA; email: sara.abedi@tamu.edu (Corresponding author)

**ABSTRACT:**

This study investigates the potential of high-speed nanoindentation in collaboration with data analytics and phase volume fractions to achieve micromechanical characterization of heterogeneous rocks. While micromechanical characterization can be performed using mechanical testing alone, integrating chemical analysis—such as elemental mapping techniques—provides essential phase identification. This enables more accurate interpretation of phase-specific mechanical properties. However, incorporating chemical analysis increases the complexity of the process. Hence, this study proposes data-driven micromechanical characterization and mapping of heterogeneous rocks based primarily on mechanical data and limited dependence on chemical analysis. In this study, Mancos shale rock is analyzed using high-speed nanoindentation with high spatial resolution to determine the mechanical properties at the microscale. Subsequently, a suite of unsupervised statistical learning techniques, such as Uniform Manifold Approximation and Projection (UMAP) with k-means Clustering, Gaussian Mixture Model (GMM), Dirichlet Process Mixture Model (DPMM), and Density-Based Spatial Clustering of Applications with Noise (DBSCAN), are applied to the nanoindentation data. Additionally, an automated image processing and segmentation technique was developed and tested. The results from each technique have been systematically compared against the conventional chemo-mechanical approach using two metrics: weighted error, which evaluates how well cluster sizes match actual phase sizes, and spatial error, which quantifies the accuracy of mineral phase assignment at each indentation point. Based on the results, UMAP with k-means clustering is the most appropriate technique (with relatively low weighted and spatial errors of 13.40% and 30.40% for Test 2 grid and 26.74% and 49.20% for Test 3 grid) for micromechanical characterization and mapping of heterogeneous rocks, while DBSCAN, DPMM, and image segmentation techniques are more suitable as secondary approaches for cross-checking results from UMAP with k-means. Moreover, a comparison of the application of SEM-EDS image volume fractions and chemo-mechanical grid volume fractions for the phase volume fractions has been performed to derive the influence of chemical analysis on the results. This study demonstrates the capability of high-speed nanoindentation combined with machine learning techniques for micromechanical characterization with reduced analytical complexity and improved workflow efficiency.

## 1 INTRODUCTION:

Assessment of variations in the mechanical properties of heterogeneous materials is essential for gaining a comprehensive understanding of their structural response and durability under varying conditions. This is particularly important for geomaterials such as shale and other sedimentary rocks, which can undergo significant compositional and microstructural changes when exposed to harsh subsurface environments—such as those encountered in $CO_2$ sequestration, geothermal operations, or enhanced oil recovery (Prakash et al., 2022, 2024). In such contexts, micromechanical characterization and mapping are essential for evaluating how these changes impact the mechanical integrity and long-term stability of the rock (Foley et al., 2012; Zhang et al., 2023). To probe these localized mechanical variations and better understand the effects of environmental exposure on rock integrity, high-resolution testing techniques are required. One of the key testing techniques to perform mechanical characterization of micro- and nanoscale materials, ranging from engineering to biological applications, is nanoindentation (Oliver & Pharr, 2004; Rossi et al., 2023; Vranjes-Wessely et al., 2021). Researchers have utilized nanoindentation testing for pore-scale mechanical analysis of cementitious materials and heterogeneous geomaterials such as fine-grained and organic-rich shale rocks (Abedi, Slim, & Ulm, 2016; Vranjes-Wessely et al., 2021; Yang et al., 2020). Nanoindentation testing provides localized measurements of mechanical properties such as hardness and elastic modulus, offering insights into the microscale behavior of heterogeneous rocks. The procedure entails applying a compressive load via an indenter tip while simultaneously monitoring and recording the penetration depth (Sobhbidari & Hu, 2021). Since a comprehensive study of heterogeneous rocks may require a large number of indentation points, high-speed nanoindentation mapping (HSNM) has emerged as a suitable and efficient approach for such materials. Compared to traditional nanoindentation, HSNM offers several advantages (Rossi et al., 2023; Vranjes-Wessely et al., 2021).

Significant research work has been published recently on the use of HSNM to investigate the mechanical properties of materials such as cement (Němeček et al., 2020; Sebastiani et al., 2016), wood (Xu et al., 2022), minerals (Beirau et al., 2019; Vranjes-Wessely et al., 2021), metals (Texier et al., 2024), 3D printed metals (Z. Liu et al., 2021), and advanced composites (Roa et al., 2018). Among these, a study of particular importance is Němeček et al. (2020), which employed HSNM for the assessment of local mechanical properties of three different heterogeneous cementitious samples, namely cement paste, fly ash, and slag blended cement pastes. The results were quantitatively and qualitatively comparable to the quasi-static indentation results, demonstrating the effectiveness and accuracy of this technique. In addition, this technique was capable of identifying the mechanical vulnerabilities of certain phases under high strain rate conditions (Němeček et al., 2020; Rossi et al., 2023). Another relevant study is Beirau et al. (2019), which utilized HSNM to investigate the effects of radiation damage on the mechanical properties of key minerals such as Zircon, which is significant for nuclear waste management applications (Beirau et al., 2019). However, the application of HSNM for micro-mechanical characterization and mapping of heterogeneous geomaterials, such as shale rocks, remains largely underexplored compared to other materials. One of the few studies applying HSNM to shale rocks is by Vranjes-Wessely et al. (2021), which investigated the mechanical properties of organic matter in over-mature shale using high-speed nanoindentation combined with optical and high-resolution imaging. The study employed k-means clustering to process the data and revealed significant property variation due to heterogeneity in the organic matter. The study also found that surrounding minerals enhance stiffness, whereas pores and cracks reduce it, and that higher thermal maturity leads to a decrease in elastic modulus (Rossi et al., 2023; Vranjes-Wessely et al., 2021). Overall, the work underscored the utility of HSNM in capturing micromechanical variability within complex shale systems.

Recent approaches to micromechanical characterization of heterogeneous materials highlight the need to combine mechanical testing with complementary chemical analyses to fully understand the behavior of their constituent phases. Mechanical properties are typically obtained through nanoindentation testing, while chemical analysis—such as X-Ray Diffraction (XRD) and scanning electron microscopy coupled with energy dispersive X-ray spectroscopy (SEM-EDS)—is employed to identify and evaluate the mineralogical composition and spatial distribution of phases within the material. However, chemical analysis of heterogeneous materials is often a time-consuming process. Apart from mechanical and chemical analyses, an appropriate data analysis and interpretation tool is also required for chemo-mechanical phase analysis and recognition. Currently, multiple interpretation techniques are commercially available that can be applied for data analysis and interpretation, like statistically aided data analysis techniques, and artificial intelligence (AI) aided algorithms, such as deep learning techniques (Rossi et al., 2023). Nonetheless, research evaluating the robustness and accuracy of these techniques to appropriately characterize and map micromechanical properties is limited. One such study, which assesses the capability of some of these techniques for micromechanical characterization of heterogeneous materials, was conducted by Zhang et al. (2023). This research investigated the micromechanical properties of hydrated cement paste, leveraging the nanoindentation results of 4000 indentation points. This study conducted the data analysis utilizing three different statistical data mining techniques, namely the Gaussian Mixture Model (GMM) with maximum likelihood evaluation (MLE) algorithm, probability distribution function (PDF), and cumulative distribution function (CDF) with least square estimate (LSE) algorithm. Comparison of the three techniques demonstrated the superiority of GMM over the other two techniques for indentation data analysis of hydrated cement paste. The study concluded that about 800 to 1000 indentation points are required for appropriate micromechanical characterization and phase proportions of hydrated cement paste (Zhang et al., 2023). A limited number of studies (Vignesh et al., 2019; Vranjes-Wessely et al., 2021) have also adopted the k-means clustering algorithm for micromechanical data analysis. Given recent advances in the speed and resolution of data acquisition through high-speed nanoindentation, there is a growing need to study and evaluate more advanced data mining techniques for effective micromechanical characterization and mapping of heterogeneous rocks. An additional consideration is whether such data—and the effective fusion of its derived properties—can reduce reliance on detailed chemical analyses.

Building on this motivation, the present study aims to advance the characterization of micromechanical heterogeneity in materials such as shale rocks by integrating high-speed nanoindentation data with advanced data mining techniques. This integrated approach enables more precise quantification and spatial mapping of mechanical property variations, supporting the development of scalable methodologies for characterizing complex heterogeneous materials. Such characterization is particularly relevant to energy geomechanics and subsurface engineering, where accurate modeling, design, and risk assessment rely on a detailed understanding of the underlying micromechanical phenomena that govern the behavior and stability of heterogeneous materials under realistic environmental conditions. To explore this potential, the study applies several unsupervised statistical learning methods along with a novel automated image processing and segmentation technique to mechanical property data—specifically, hardness and elastic modulus—obtained from high-speed nanoindentation tests on Mancos shale. Volume fraction information for the different phases is also incorporated to enhance clustering and interpretation. The performance of each technique is evaluated through comparison with chemo-mechanical mappings derived from coupled SEM-EDS and nanoindentation data. Furthermore, the influence of phase volume fractions on the quality of mechanical mapping is assessed by comparing SEM-EDS-based and chemo-mechanical-based (coupled SEM-EDS with indentation) volume estimates.

Ultimately, this work seeks to establish a consistent correlation between mechanical response and phase composition, offering a data-driven framework for micromechanical characterization with reduced experimental and analytical complexity. By reducing reliance on detailed chemical analysis, the proposed approach has the potential to reduce analytical complexity and support broader application in heterogeneous geomaterials research.

## 2 RESEARCH SIGNIFICANCE:

Despite significant progress in micromechanical characterization of geomaterials, most existing approaches rely on combining nanoindentation with detailed chemical or compositional analysis to identify and interpret material phases. While effective, such approaches are often time-intensive and difficult to scale when high-resolution spatial characterization is required. Moreover, the interpretation of nanoindentation data remains largely dependent on external compositional information rather than being inferred directly from the mechanical response. At the same time, recent advances in high-speed nanoindentation enable rapid acquisition of large datasets; however, there remains a lack of systematic frameworks for extracting phase-level information directly from mechanical measurements using data-driven approaches. This study addresses this gap by developing a mechanics-driven and data-centric framework that integrates high-speed nanoindentation with unsupervised data analytics to infer phase behavior primarily from mechanical responses. The approach further incorporates automated segmentation of mechanical property fields to map micromechanical heterogeneity and is evaluated against independent chemo-mechanical characterization to assess its reliability. In addition, the influence of phase volume fractions on clustering performance and phase identification is examined. By reducing reliance on detailed chemical analysis, this framework supports micromechanical characterization of heterogeneous geomaterials with reduced analytical complexity.

## 3 MATERIALS AND METHODS:

The overall framework of this study is illustrated in Fig. 1. The rock specimens were obtained from the Mancos Shale Formation. Their composition, determined by XRD, summarized in **Table 1** (Banu et al., 2025; Mahgoub et al., 2025; Nguene, 2019). Phase volume fractions from SEM-EDS images and modulus and hardness from nanoindentation are used as inputs for machine learning techniques to characterize and spatially map phase-specific mechanical properties across the shale surface. Details of each method are presented in Section 3.4.

While limited chemical information has been used in this study such as XRD (for number of phases) and SEM-EDS (for phase volume fractions). It is incorporated only as global constraints to maintain physically realistic phase proportions. In this study, the XRD results serve as a reference for the expected number of mineral phases, while clustering metrics (e.g., SSE, Silhouette Score, CHI, and DBI) are used to evaluate and determine the optimal number of clusters. These results are then validated against the XRD-based phase information. The chemical inputs guide cluster number and sizes or assist in interpreting cluster assignments but do not directly influence per-indentation-point classification. Therefore, the approach remains largely data-driven, with mechanical properties serving as the primary basis for phase differentiation and only limited dependence on chemical analysis across the clustering and image processing–segmentation techniques.

### 3.1 Specimen Preparation:

Accurate results from surface-sensitive tests such as high-speed nanoindentation and SEM-EDS require flat surfaces parallel to the specimen bottom. Accordingly, shale specimens were cut, ground, and trimmed into cube-shaped samples with side lengths of 10–20 mm. Initial coarse polishing was carried out with a 400-grit abrasive disc to expose surface textures. The surfaces were then progressively refined using aluminum oxide abrasive discs (FibrMet, Buehler) with particle sizes of 12, 9, 3, 1, and 0.30 µm. Between each polishing step, the specimens were ultrasonicated in an n-decane solution, which was selected for its non-reactivity with rock. For additional details on the specimen preparation procedure, the reader is referred to (Abedi, Slim, Hofmann, et al., 2016; Martogi & Abedi, 2020; Sharma et al., 2019).

### 3.2 High Speed Nanoindentation Testing:

Nanoindentation testing is a powerful technique for evaluating the pore-scale mechanical behavior of heterogeneous materials. It involves impinging the indenter tip—typically of known geometry, such as a diamond Berkovich tip—into the material surface to extract localized mechanical properties. In this study, high-speed nanoindentation mapping (HSNM) was conducted using the NanoBlitz 3D nanoindenter (iMicro, KLA Tencor, USA) (Banu et al., 2025). The HSNM method is particularly valuable as it enables thousands of indents to be performed rapidly, facilitating the generation of high-resolution spatial maps of indentation modulus and hardness over large surface areas. Each indent requires approximately one second, including the time for surface approach, loading, unloading, tip withdrawal, and repositioning of the specimen for the next measurement. To determine the optimal and most efficient indentation depth and spacing, multiple trial-and-error experiments were performed on the specimens. An indentation depth greater than 50 nm is recommended to obtain the inherent modulus and hardness. Furthermore, the indentation spacing to indentation depth ratio is highly recommended to be greater than 10 to prevent any overlap between the probed volumes (Z. Liu et al., 2021; Phani & Oliver, 2019). In this study, a linear loading of 4.8 mN along with an unloading phase were applied to the specimen during the indentation test (Abedi, Slim, Hofmann, et al., 2016). Four indentation grids were performed on the shale rock specimen, following the calibration of the area function on fused silica specimens and the load frame compliance correction (Vignesh et al., 2019). The indentation spacing for each grid was about 10 µm, and each grid was composed of 2000 indentation points, covering a surface area of approximately 500 × 400 µm². For this study, the mechanical data from two indentation grids, namely, Test 2 (second indentation grid) and Test 3 (third indentation grid) grids out of the four tested grids, will be utilized and investigated. The mechanical properties were obtained from the load–displacement (*P-h*) curve plotted by the nanoindenter by applying the continuum scale model to derive the indentation modulus curves generated by the nanoindenter, with the indentation modulus (*M*) and hardness (*H*) derived using the continuum-scale model. These parameters were calculated following the Oliver and Pharr method (Banu et al., 2025; Oliver & Pharr, 1992), as expressed in Eqs. (1) and (2) (Abedi, Slim, Hofmann, et al., 2016):

$$M = \frac{\sqrt{\pi}}{2}\frac{S}{\sqrt{A_c}} = \frac{E}{1-\nu^2} \quad (1)$$

$$H = \frac{P}{A_c} \quad (2)$$

where $A_c$ is the projected contact area between the indenter tip and the specimen surface, estimated as a function of the measured maximum indentation depth, $P$ represents the measured maximum indentation load, $S = \frac{dP}{dh}$ denotes the contact stiffness, that is, the initial slope of the unloading segment of the *P-h* curve, E represents the Young's modulus, and $\nu$ is the Poisson's ratio.

Notably, no corrections to the Oliver–Pharr method were required, as minimal pile-up was observed during high-speed nanoindentation testing.

### 3.3 Determining the Optimal Number of Mineral Phases:

Most unsupervised clustering algorithms require the number of clusters (i.e., mineral phases) to be specified beforehand. Hence, to identify the appropriate number of mineral phases for clustering, several internal validation metrics were employed, including the Sum Square Error (SSE) or Elbow method (Marutho et al., 2018; Shi et al., 2021), Silhouette score (Shi et al., 2021), Calinski–Harabasz Index (CHI) (Wang & Xu, 2019), and Davies–Bouldin Index (DBI) (Mughnyanti et al., 2020; Wani & Riyaz, 2017). These methods assess clustering quality based on criteria such as intra-cluster cohesion, inter-cluster separation, and overall variance. Implementation was carried out using the scikit-learn library in Python. The combination of these metrics helped ensure a robust estimate of the optimal number of phases, which was then used in subsequent unsupervised clustering analyses.

The SSE or Elbow method is a well-established clustering evaluation metric. It is defined as the sum of the squared Euclidean distances between each data point and its corresponding cluster centroid. The optimum number of clusters occur at the "elbow" point, where SSE shows a sharp decrease and the curve forms a distinct angle. For further detailed information on the SSE or Elbow method evaluation metric and its formulation, readers are referred to (Marutho et al., 2018).

The silhouette score is calculated using Eq. (3) (Shi et al., 2021; Wang & Xu, 2019):

$$Silhouette\ Score = \frac{1}{n}\sum_{i=1}^{n}\left(\frac{b(i)-a(i)}{\max\{a(i),b(i)\}}\right) \tag{3}$$

where $a(i)$ is the mean intra-cluster distance (average distance of sample $i$ to other samples in the cluster) and $b(i)$ is the mean closest cluster distance (minimum distance of the sample $i$ to the other clusters). It ranges from -1 to 1, with values closer to 1 indicating optimal clustering.

The CHI index value was determined based on the following equations (4-6):

$$CHI(K) = \frac{B(K)(N-K)}{W(K)(K-1)} \tag{4}$$

$$B(K) = \left(\sum_{k=1}^{K} a_k \left|\left|\overline{x_k} - \overline{x}\right|\right|^2\right) \tag{5}$$

$$W(K) = \left(\sum_{k=1}^{K}\sum_{c(j)=k}\left|\left|x_j - \overline{x_k}\right|\right|^2\right) \tag{6}$$

where, $K$ stands for the number of clusters, $B(K)$ is the inter-cluster divergence (or inter-cluster covariance), $W(K)$ represents the intra-cluster divergence (or intra-cluster covariance), $N$ denotes the number of samples, $a_k$ represents the number of data points in cluster $K$, $\overline{x_k}$ is the centroid of the k[th] cluster, $\overline{x}$ denotes the global centroid and $x_j$ is the j[th] data point. Higher values of CHI indicate better clustering (Wang & Xu, 2019).

Finally, Eq. (7) was used to compute the DBI for $K$ number of clusters:

$$DBI(K) = \frac{1}{K}\left(\sum_{i=1}^{K} max_{j\neq i} \frac{(\Delta C_i + \Delta C_j)}{\delta(C_i, C_j)}\right) \tag{7}$$

where, $\Delta C_i$ and $\Delta C_j$ represent the intra-cluster distances, estimated as the average distance of all the cluster elements $C_i$ and $C_j$ to their respective cluster centroids, and $\delta(C_i, C_j)$ represents the distance between cluster centroids ($C_i$ and $C_j$). Lower values of DBI indicate better clustering (Mughnyanti et al., 2020; Wani & Riyaz, 2017).

### 3.4 ML Techniques for Micromechanical Characterization and Mapping of Heterogeneous Rocks:

Different clustering and segmentation techniques vary in assumptions, sensitivity, and ability to capture underlying structure; thus, systematic evaluation is needed to identify the most suitable approach for micromechanical phase differentiation.

This study implemented four clustering techniques—UMAP with k-means, GMM, DPMM, and DBSCAN—alongside an image processing and segmentation method, using standardized indentation modulus ($M$) and hardness ($H$) data. The effectiveness of each clustering technique in micromechanical characterization was evaluated using two error metrics: weighted error, which assesses how well cluster sizes match actual phase proportions, and spatial error, which quantifies the accuracy of phase assignment at each indentation point. Lower values of these errors indicate optimal accuracy in characterizing and spatially mapping micromechanical phases. Only standardization was performed on these indentation features ($M$ and $H$) because the dataset exhibited minimal outliers and high consistency across the indentation grid, making additional data filtering unnecessary. Although relying solely on $M$ and $H$ can limit phase separability when different constituents exhibit overlapping mechanical properties, this limitation is inherent to micromechanical characterization of highly heterogeneous materials and can affect clustering performance.

The choice of $M$ and $H$ is intentional and aligned with the study's objective. These directly measured parameters provide a consistent, physically interpretable basis for large-scale, high-throughput characterization. The goal is to assess whether high-speed nanoindentation, combined with data-driven clustering and phase volume constraints, can produce meaningful micromechanical maps without extensive chemical characterization or additional experimental inputs. While adding derived features (e.g., ratios, composite indices, or inverse-derived parameters) could improve phase separation, such approaches introduce assumptions, model dependencies, and complexity, shifting the framework away from the intended streamlined, mechanics-driven methodology suitable for scalable application.

The general workflow followed for all the adopted clustering techniques, excluding image processing and segmentation approach, is illustrated in Fig. 2. Phase volume fractions were incorporated as global prior constraints to maintain physically realistic phase distributions. The number of clusters was determined using domain knowledge and validated heuristically (Elbow and Silhouette methods). Results were visualized via 2D spatial maps and modulus–hardness scatter plots, and compared with chemo-mechanical SEM-EDS results to compute weighted and spatial errors.

Convergence and reproducibility were evaluated using Silhouette scores and Adjusted Rand Index (ARI) across multiple random seeds. For UMAP with k-means, multiple runs were performed per hyperparameter combination, with Silhouette scores and ARI used to assess clustering consistency. The ARI quantifies the similarity between the resulting cluster labels and a reference (ground truth) model, with higher values indicating stronger agreement (Chacón & Rastrojo, 2023). It was computed using Eq. (8) (Das & Biswas, 2023):

$$ARI\ (C^r, C^m) = \frac{Index - Expected\ Index}{Maximum\ Index - Expected\ Index} = \frac{\sum_{ij}\binom{n_{ij}}{2} - \frac{\left[\sum_i \binom{a_i}{2}\sum_j \binom{b_j}{2}\right]}{\binom{n}{2}}}{\frac{1}{2}\left[\sum_i \binom{a_i}{2} + \sum_j \binom{b_j}{2}\right] - \frac{\left[\sum_i \binom{a_i}{2}\sum_j \binom{b_j}{2}\right]}{\binom{n}{2}}} \tag{8}$$

where, $C^r$ and $C^m$ represent the reference (or true) model clusters and current model clusters respectively, $n_{ij}$ is the number of nodes in $C_i^m$ and $C_j^r$ clustering, $a_i$ is the sum of all the $n_{ij}$ corresponding to any $C_j^r$ of $C^r$ and all the $C_i^m$ of $C^m$, whereas $b_j$ is the sum of all the $n_{ij}$ corresponding to any $C_i^m$ of $C^m$ and all the $C_j^r$ of $C^r$. ARI values range from -1 to 1, where 1 indicates perfect agreement with the reference model (Das & Biswas, 2023). Convergence was checked via ARI for the other techniques as well.

### 3.4.1 UMAP with k-means Clustering

Uniform Manifold Approximation and Projection (UMAP) is a nonlinear dimensionality reduction technique that maps high-dimensional data to a lower-dimensional space while preserving both local and global structures (Dorrity et al., 2020; Healy & McInnes, 2024). This technique was selected for its ability to reduce dimensionality while maintaining local structure, enabling clear phase separation. Its performance depends on hyperparameters such as the number of neighbors (n_neighbors), minimum distance (min_dist), embedding dimensionality (n_components), and the distance metric (Euclidean or Manhattan) (Sahani, 2020). Standardized indentation modulus and hardness values, along with phase volume fractions, were used as input features, with phase fractions supporting the definition of target cluster sizes. Following dimensionality reduction, k-means clustering grouped data points with similar mechanical properties (Rossi et al., 2023). The number of clusters was set to match the distinct phases in the rock specimen, and a two-stage clustering approach was used to separate major and minor phases. A limitation in the context of this study is that UMAP's effectiveness can be sensitive to hyperparameter selection and may occasionally distort global relationships, potentially affecting the representation of minor phases when volume fractions are very small.

### 3.4.2 GMM Clustering

Another clustering technique evaluated is the Gaussian Mixture Model (GMM), a probabilistic method based on maximum likelihood estimation. GMM was included as a classical probabilistic baseline for comparison with the non-parametric clustering approaches, allowing assessment of how a distribution-based method performs within the same micromechanical feature space. In GMM, each cluster is modeled as a Gaussian distribution in the indentation modulus–hardness feature space, enabling probabilistic assignment of points to phases. The model estimates the weight, mean, and covariance of each component and assigns points to clusters based on posterior probabilities. Parameters are determined via the Expectation–Maximization (EM) algorithm, iteratively updating cluster membership probabilities until convergence (Zhang et al., 2023). Phase volume fractions were used as initial mixture weights to guide convergence toward realistic phase proportions, and multiple random seeds were tested to assess solution stability. A limitation of GMM is that it assumes clusters are Gaussian, which may not perfectly reflect the actual distribution of mechanical properties, especially for minor phases with skewed or non-Gaussian behavior. Consequently, small-volume or overlapping phases may be less accurately resolved.

### 3.4.3 Dirichlet Process Mixture Model (DPMM) Clustering

The Dirichlet Process Mixture Model (DPMM) is a nonparametric Bayesian clustering approach that infers the number of clusters directly from the data, without requiring it to be specified a priori (Sun & Zheng, 2020). This is achieved via the Dirichlet Process (DP), which uses a concentration parameter to control the likelihood of forming new clusters (Stratton et al., 2024) and a base distribution as a prior for cluster-specific parameters. DPMM was selected because it relaxes the fixed-cluster assumption, allows probabilistic assignment of data points, and better handles variable cluster sizes. In the generative framework, each observation is drawn from a mixture distribution governed by DP-distributed latent parameters, so multiple observations sharing the same parameters naturally form clusters. Phase volume fractions were incorporated to inform the weight concentration prior, guiding the model toward realistic cluster proportions, and multiple random seeds were tested to assess robustness and stability (Stratton et al., 2024). A limitation in this study is that, although DPMM can flexibly assign points to clusters, it may still be influenced by the initial prior settings and hyperparameters, particularly for very small or overlapping phases. Consequently, minor phases may sometimes be underrepresented or merged if their mechanical properties closely resemble neighboring clusters.

#### 3.4.4 DBSCAN Clustering

Density-Based Spatial Clustering of Applications with Noise (DBSCAN) is a widely used clustering algorithm that identifies clusters based on data density, making it effective for irregular cluster shapes and noisy datasets. It has been applied in various geoscience contexts, such as wireline logs (Ali et al., 2023), sonic logs (He et al., 2019), and drilling data (Hansen & Aarset, 2024). DBSCAN requires two parameters: the maximum neighborhood distance ($\varepsilon$) and the minimum number of points (min_samples) to define a dense region (Mustafa et al., 2024). Points meeting these criteria form clusters, while points that do not are labeled as outliers. Overly small $\varepsilon$ values may create excessive outliers, whereas overly large $\varepsilon$ values may merge distinct clusters; min_samples selection also depends on dataset size and dimensionality (Mustafa et al., 2024). DBSCAN was applied across a wide range of $\varepsilon$ and min_samples values to assess clustering behavior. It was selected for this study due to its ability to identify clusters without assuming any specific shape, unlike Gaussian-based methods. A limitation for the current indentation data is that DBSCAN does not incorporate prior information, such as phase volume fractions, during clustering. Consequently, small or sparsely represented phases may be misclassified or labeled as noise, particularly when mechanical properties of neighboring phases are similar, a scenario common in heterogeneous rocks like shale. Phase volume fractions were therefore used post hoc to validate cluster correspondence with measured phase abundances.

#### 3.4.5 Image Processing and Segmentation Technique

This method was selected for its ability to leverage spatial continuity and gradient-based information, enabling direct detection of mechanically distinct regions and improved delineation of phase boundaries. Indentation modulus and hardness values were normalized to a 0–255 scale and mapped onto a two-dimensional spatial grid to preserve the specimen's internal heterogeneity. Classification bins were then defined based on these normalized values, with thresholds corresponding to the known mechanical ranges of each mineral phase. These thresholds mapped grayscale pixel intensities to distinct bins, ensuring each segmented region represented a specific mineral phase. The same mechanical thresholds were applied consistently across all images, allowing physically meaningful phase assignments while maintaining spatial heterogeneity. The composite image was converted to grayscale (Tang, 2020), and Sobel filters were applied to compute gradient magnitudes, highlighting mechanical boundaries. Contour mapping was then used to segment mechanically distinct zones, with contour levels determined from the cumulative distribution of gradient magnitudes and calibrated to match known phase volume fractions, and the resulting zones were assigned to mineral phases based on established mechanical thresholds. Phase volume fractions were incorporated as global pixel-percentage constraints during threshold-based segmentation to ensure physically realistic phase abundances, rather than as per-pixel classification features. A limitation of this technique is that it relies on global phase constraints and predefined thresholds, which may reduce sensitivity to local mechanical variations. It may also be less effective in regions with very sparse or overlapping phases, where boundary detection becomes challenging.

### 3.5 Mineralogical and Chemical Analysis of the Rock Specimen:

The phase volume fractions used in the clustering techniques were estimated using the chemical analysis of the rock specimen. Firstly, SEM-EDS was conducted on the shale sample over the indentation grid area to determine its elemental composition. SEM imaging qualitatively captured the 2D surface morphology, while EDS provided qualitative information on the elemental distribution. Unlike optical imaging, SEM employs a focused electron beam to generate high-resolution images of the specimen surface (Y. Liu et al., 2022). To prepare the rock specimen for SEM imaging, the polished surface was coated with a thin layer of carbon, preventing any heavy charging effects during SEM imaging. The acceleration voltage for this study, was varied between 10 to 20 keV for the indented area, and the system vacuum pressure used was about 10-4 Pa. SEMs are capable of attaining a spot size of about 10 nm (Prakash, 2022). To ensure accurate alignment of the SEM-EDS maps with the indented region, a shallow slit mark was engraved on the specimen surface after the indentation tests, slightly away from the indented area to avoid affecting the mechanical measurements. This mark served as a positional reference for locating the indentation grid, since the indentation points themselves are not directly visible in the SEM-EDS images. Following this, the SEM images of the indented grid areas were captured. Next, qualitative EDS was performed to identify and quantify the chemical elements present in the sample (Y. Liu et al., 2022). These SEM-EDS images were further processed using multispectral image analysis to extract mineralogical information and phase volume fractions for each grid area (Prakash et al., 2021) , and, combined with XRD phase identification, provided reliable SEM-EDS–derived mineralogical maps of the indented regions to support mineralogical interpretation rather than quantitative compositional analysis. Finally, the chemical data were manually spatially coupled with the nanoindentation results to determine the representative mineral phase under each indentation point (as shown in Fig. 3). This approach enabled correlation of mechanical properties—modulus and hardness—with specific mineral phases, and supported the generation of chemo-mechanical maps for the tested rock specimen. For further details regarding the EDS results, the readers are referred to (Mahgoub et al., 2025). Figure 3 presents the multispectral image (the grayscale background image) of Test 2 grid area, which provides the SEM-EDS image volume fractions for each mineral phase. The indentation grid in the overlaid image appears tilted because the sample was slightly angled during indentation test relative to the SEM-EDS imaging. To ensure accurate alignment, the grid was rotated to match the actual indented locations.

### 3.6 Error Metrics for Evaluating Clustering Performance:

As mentioned, to select the most optimal results from each clustering technique, two error measures were computed, namely weighted and spatial errors. The results from each adopted technique (clustering and image processing and segmentation) were then compared based on these error values to identify the most appropriate method for accurate micromechanical characterization and spatial mapping. The weighted error is the summation of number of data points in each cluster ($n_i$) multiplied by the error (%) between the clustered/computed ($y_i$) and true/expected phase volume fractions ($\hat{y}_i$), divided by the total number of points ($N$). The true phase volume fractions were obtained from the ground truth model prepared based on the chemo-mechanical results. The equation for weighted error is provided in Eq. (9).

$$Weighted\ Error = \frac{\sum n_i * Error\ (\%)}{N} = \frac{\sum n_i * \left(\frac{(y_i - \hat{y}_i)}{\hat{y}_i} \times 100\right)}{N} \tag{9}$$

When true phase volume fraction of a phase was zero, absolute error $(y_i - \hat{y}_i)$ was used for error (%). For the spatial error, the 2D spatial map generated from each clustering technique were compared against the ground truth model. Equation (10) presents the spatial error formula, where $n_{accurate}$ represents the number of accurately classified data points and $N$ is the total number of data points.

$$Spatial\ Error = 100 - \left(\frac{\sum n_{accurate}}{N}\right) * 100 \tag{10}$$

## 4 RESULTS AND DISCUSSION:

### 4.1 Evaluation of the Optimal Number of Mineral Phases:

The optimal number of clusters (interpreted as distinct mineral phases) for the indentation grids obtained from HSNM was estimated using four evaluation metrics: SSE, Silhouette score, CHI, and DBI. The results for Test 2 and Test 3 indentation grids, based on these methods, are provided in Fig. 4 and reported in

**Table 2**. The number of clusters was incrementally increased (from 2 to 6) until convergence toward an optimal value was observed for each metric. The range of clusters (2–6) was selected based on both mineralogical considerations and literature-reported phase compositions of shale materials. Previous studies (Banu et al., 2025; Mahgoub et al., 2025; Prakash et al., 2024) have shown that quartz-rich shales typically consist of a limited number of dominant mineral phases—commonly clay minerals, quartz, feldspar, calcite, dolomite, and pyrite—resulting in approximately 4–6 mechanically distinguishable phases. However, depending on spatial resolution and mechanical overlap, some phases may merge into composite clusters, particularly at lower cluster numbers. Accordingly, the lower bound (2 clusters) was chosen to evaluate whether the data naturally separate into broad mechanical groupings (e.g., stiff vs. compliant phases), while the upper bound (6 clusters) corresponds to the maximum number of independently identified mineral phases in the studied material based on the XRD results and established shale characterization literature. Thus, the selected range ensures both methodological rigor and physical relevance to the known mineralogical complexity of the material.

For both Test 2 and Test 3 indentation grids, sum of squared errors (SSE) shown in Fig. 4(a) and reported in **Table 2** shows that it decreases smoothly with increasing clusters, reaching the lowest value at six clusters; however, the absence of a clear elbow limits its ability to definitively determine the optimal cluster number (Shi et al., 2021). In contrast, the Silhouette Score (Fig. 4(b)) provides stronger guidance. For Test 2, the score increases up to five clusters and decreases slightly at six, though the reduction is marginal. For Test 3, the score peaks at three clusters but shows a minor decline thereafter. Considering the minimal change and the chemo-mechanical analysis indicating six distinct mineral phases, six clusters were selected. This choice is further supported by the Calinski–Harabasz Index (CHI) (Fig. 4(c)), which reaches its highest value at six clusters for both tests. The Davies–Bouldin Index (DBI) (Fig. 4(d)), decreases up to five clusters for Test 2 and four clusters for Test 3, followed by a slight increase at six clusters, indicating marginal over-partitioning at higher cluster numbers.

Overall, these clustering metrics indicated a range of plausible cluster numbers—five to six for Test 2 and fewer for Test 3—reflecting material heterogeneity and phase dominance rather than limitations in the analysis. For instance, pyrite in Test 2 has a minor volume fraction (~0%), reducing its statistical representation and biasing some metrics toward fewer clusters, while Test 3 is dominated by three major phases (~84% of the indented region), naturally favoring fewer clusters. The selection of six clusters was therefore guided by independent XRD data, confirming six primary mineral phases in the Mancos shale. This hybrid framework integrates data-driven clustering with physically grounded constraints, ensuring that the resulting clusters correspond meaningfully to known mineralogy while capturing the mechanical response.

### 4.2 Micromechanical Mapping Results from Clustering and Image Segmentation:

This subsection presents the results of micromechanical characterization and mineralogical mapping using high-speed nanoindentation data, mineral phase volume fractions, and the techniques described in Section 3.4. Four clustering methods—UMAP with k-means, GMM, DPMM, and DBSCAN—were applied alongside an image processing and segmentation technique to the Test 2 grid of the HSNM dataset to evaluate their effectiveness. For all techniques, the number of clusters was fixed at six, based on the optimal cluster count determined earlier.

To assess the sensitivity of clustering performance to input volume fractions, two types of mineral phase volume fractions were tested: (1) SEM-EDS image volume fractions, derived from multispectral image analysis of SEM-EDS data; and (2) chemo-mechanical grid volume fractions, obtained by coupling high-speed nanoindentation results with SEM-EDS image data. **Table 3** summarizes both sets of volume fractions for the six mineral phases identified in the Test 2 grid. The overall error between the SEM-EDS image and chemo-mechanical grid volume fractions was found to be approximately 11.41% (see **Table 3**), indicating reasonable agreement with minor discrepancies likely arising from spatial resolution or phase boundary definitions.

The chemo-mechanical results along with their modulus and hardness of Test 2 grid are presented in Fig. 5 and summarized in **Table 4**. These results serve as the reference for validating the outcomes of the data-driven clustering techniques. Due to the small volume fraction and low significance of the seventh mineral phase labeled "Others," which represents minor constituents co-occurring with major phases (provided in **Table 1**) and not identified as a distinct phase in the XRD mineralogy, it was excluded from the clustering analysis and its volume redistributed equally among the six major mineral phases to ensure consistency across clustering results. In Figs. 5-12, the mean modulus and hardness values for each respective phase are also provided in the color bar.

Due to the fine-scale heterogeneity of shale, it is challenging to ensure that every nanoindentation is placed entirely within a single mineral phase. This limitation is well recognized in the literature, and nanoindentation results in geomaterials are often interpreted in terms of phase-dominant responses (e.g., "quartz-rich" or "clay-rich") rather than strictly single-phase measurements. A similar approach is adopted in this study. Indents located near phase boundaries contribute to the natural scatter observed in the measured modulus and hardness values. Importantly, the presence of a limited number of boundary-affected indents does not bias the overall interpretation, as the analysis focuses on dominant trends and representative ranges of mechanical properties. The majority of indents are situated within phase-dominant regions; therefore, the reported results are considered representative of the mechanical response of the corresponding mineral phases.

#### 4.2.1 Results from Data-Driven Clustering Techniques

This subsection presents a comparative evaluation of four data-driven clustering techniques—UMAP with k-means, GMM, DPMM, and DBSCAN—for micromechanical characterization and mapping of the Test 2 grid using HSNM data. Each technique was applied using six fixed clusters and evaluated with both SEM-EDS image-derived and chemo-mechanical grid-based volume fractions. Results were assessed through 2D spatial maps, modulus–hardness scatter plots, and tabulated metrics (Figs. 6-9; **Tables 5–8**), with validation against the chemo-mechanical reference results (Fig. 5; **Table 4**).

UMAP with k-means: UMAP with k-means yielded the most accurate and consistent results (Fig. 6; **Table 5**). Using SEM-EDS image volume fractions, the method effectively identified clay and quartz phases, with partial success for feldspar. Weighted and spatial errors were 13.40% and 30.40%, respectively. When chemo-mechanical grid volume fractions were used, weighted error dropped to 3.00%, although spatial error remained similar (35.15%). The modulus and hardness values of major phases aligned well with reference data (**Table 4**), and ARI values remaining consistent or exceeding 0.90 across multiple runs confirmed strong reproducibility. These results demonstrate that UMAP with k-means is reliable for characterizing dominant phases, and even the SEM-EDS variant offers acceptable performance with reduced dependency on reference maps.

GMM: GMM clustering exhibited the poorest performance, with substantial misclassification and highly inflated error metrics. As shown in Fig. 7 and **Table 6**, clusters were skewed—particularly over-assigning calcite and dolomite and under-identifying clay and quartz. Weighted and spatial errors reached 469.61% and 81.30% with SEM-EDS image volume fractions, respectively, and 521.13% and 75.95% with chemo-mechanical input. Modulus and hardness values deviated significantly from reference values, especially for quartz and pyrite. Despite high ARI values, the poor clustering accuracy renders GMM unsuitable for micromechanical phase mapping in this context. This limitation primarily arises from the GMM's inherent assumption that the underlying indentation data follow Gaussian distributions, coupled with its sensitivity to noise and outliers. Analysis of the indentation parameters (modulus and hardness) indicated non-Gaussian characteristics, including pronounced skewness and long-tailed behavior. Consequently, the Gaussian assumption intrinsic to GMM does not accurately reflect the true statistical structure of the data, reducing its effectiveness for clustering in this context.

DPMM: DPMM clustering provided moderate performance, outperforming GMM but falling short of UMAP and DBSCAN. **Table 7** and Fig. 8 show accurate identification of major phases such as clay and quartz, while minor phases like calcite, dolomite, and pyrite were inconsistently mapped. Weighted and spatial errors were 147.26% and 43.45%, respectively, regardless of the volume fraction type used—indicating that DPMM clustering is less sensitive to input phase volume fractions and more reliant on the underlying mechanical data. Although ARI values approached 1.00, suggesting high replicability, the method's limited resolution of minor phases reduces its standalone utility.

DBSCAN: DBSCAN clustering (Fig. 9; **Table 8**) achieved better overall accuracy than GMM and DPMM. Using eps = 0.14 and min_samples = 11, DBSCAN effectively identified clay and quartz phases, with minor phases like calcite and dolomite less precisely resolved. Weighted and spatial errors were 19.41% and 31.70%, respectively, and consistent across both input volume fraction types—highlighting the method's insensitivity to input variability. Notably, a high number of noise points had to be manually reassigned to quartz based on prior knowledge. While this post-processing limits DBSCAN's automation, its lower error rates and reproducibility across random seeds suggest it may serve as a useful secondary clustering method. These observations align with the expected clustering behavior in heterogeneous rocks: DBSCAN reliably captures major, well-represented phases, but minor or sparsely distributed phases may be fragmented or labeled as outliers.

#### 4.2.2 Results from Image Processing and Segmentation Technique

Image processing and segmentation results using SEM-EDS image volume fractions and chemo-mechanical grid volume fractions are shown in Fig. 10, with quantitative metrics summarized in **Table 9**. This technique directly segments mechanical property maps based on the input volume fractions, resulting in near-identical input and output fractions. Consequently, the weighted errors are very low—11.49% and 0.13% for SEM-EDS and chemo-mechanical grid inputs, respectively.

Despite the low weighted error, visual inspection of the spatial maps and modulus–hardness plots (Fig. 10) reveals significant misclassification. Most data points are incorrectly labeled as clay, including regions that should correspond to quartz and feldspar based on chemo-mechanical reference data (Fig. 5). This over-assignment leads to unrealistic modulus and hardness ranges for clay (e.g., up to 157.5 GPa and 24.5 GPa), deviating from literature and reference values (Li et al., 2022; Yang et al., 2020). Spatial errors, which better capture the agreement with chemo-mechanical mapping, were 46.60% and 48.65% for SEM-EDS and chemo-mechanical inputs, respectively (**Table 9**). These results indicate limited effectiveness in resolving minor phases and moderate misclassification among major phases. As the segmentation is highly dependent on the phase volume fractions and lacks sensitivity to spatial or mechanical distinctions, its utility is constrained unless phase distributions are highly accurate and representative. Nevertheless, this method can be considered a viable approach for the approximate mapping of dominant phases—particularly clay—due to its reliance on predefined volume fractions, which helps reduce weighted errors. The main advantage of this technique is its flexibility in adjusting phase thresholding to closely match the input phase volume fractions, often resulting in reasonably accurate results. Incorporating additional chemical indicators, such as mineral-specific intensity thresholds or accurate gray scale values, may improve the accuracy further. Given its simplicity, flexibility, and efficiency, image segmentation may serve as a secondary technique to support or validate results from more robust clustering methods.

#### 4.2.3 Summary of Comparative Observations and Recommended Approaches

Among the techniques evaluated, UMAP with k-means clustering demonstrated the most consistent and accurate micromechanical characterization and mapping performance. It produced the lowest weighted and spatial errors across all clustering methods, while yielding modulus and hardness values for dominant phases (clay and quartz) that closely matched the chemo-mechanical reference results. Since UMAP with K-means involved testing numerous hyperparameter combinations compared to the other techniques, it required relatively higher computational effort. Yet, for moderate grid sizes, such as the grids used in this study, the clustering approaches and image processing and segmentation techniques are not computationally intensive. However, for larger grids exceeding 10,000 indentation points, computational demands would increase, but the methods could still be executed efficiently on conventional computing hardware, with processing time and memory usage naturally rising with grid size.

Notably, results obtained using SEM-EDS image volume fractions were comparable to those using chemo-mechanical grid volume fractions, with only marginal differences in accuracy. This highlights the practical advantage of using SEM-EDS inputs, which offer reduced dependence on more complex and time-consuming chemo-mechanical analysis, thereby supporting more streamlined and scalable micromechanical characterization.

Using the Test 2 grid and SEM-EDS image volume fractions, the mean modulus values of clay, quartz, feldspar, calcite, dolomite, and pyrite were found to be 29.99 GPa, 77.85 GPa, 109.75 GPa, 11.85 GPa, 13.85 GPa, and 8.12 GPa, respectively. Corresponding mean hardness values were 1.40 GPa, 10.30 GPa, 17.41 GPa, 0.25 GPa, 0.28 GPa, and 0.14 GPa. These values were consistent with those from the chemo-mechanical analysis (Fig. 5; **Table 4**) for high-concentration phases, particularly clay and quartz. For low-concentration phases such as feldspar, calcite, dolomite, and pyrite, clustering accuracy was limited due to the scarcity of representative data points, highlighting a general limitation shared across all techniques.

Overall, UMAP with k-means is identified as a strong candidate for micromechanical characterization and mapping of heterogeneous rocks using HSNM data. When combined with SEM-EDS image volume fractions, it demonstrates a balance of accuracy and reproducibility within the conditions tested. Other methods, such as DPMM, DBSCAN, and direct image segmentation, may serve as secondary tools for cross-validation, while GMM are less reliable under the conditions tested.

### 4.3 Evaluation of Clustering Behavior Under Varying Grid Conditions:

To evaluate the generalizability of UMAP with k-means clustering, an independent HSNM indentation dataset (Test 3 grid), not used during the development phase of the workflow, was analyzed using both SEM-EDS image and chemo-mechanical grid volume fractions. Chemo-mechanical reference results for this grid are presented in Fig. 11 and **Table 10**. Notably, some discrepancies were observed in the chemo-mechanical analysis—particularly for quartz, feldspar, and pyrite, where the modulus and hardness values are relatively lower than expected. These differences can be attributed, to material heterogeneity and challenges in precisely aligning indentation locations with the SEM-EDS image, especially given the micro-scale nature of the analysis, where small positional uncertainties can influence phase assignment. Volume fraction discrepancies between SEM-EDS and chemo-mechanical data were more pronounced in Test 3, with an average error of 28.01% (**Table 11**), reflecting the high heterogeneity of the shale specimen and the complexity of accurately capturing spatial phase distributions. Such effects are further amplified in regions where minor phases are sparsely distributed. Clustering was performed using the same UMAP with k-means framework as for Test 2, and results are shown in Fig. 12, with quantitative metrics in **Table 12**.

Using SEM-EDS image volume fractions (Fig. 12(a)), clustering reasonably identified the dominant clay phase but showed poor spatial placement of minor phases due to their sparse distribution. Mean modulus and hardness values for clay and quartz were consistent with the chemo-mechanical reference, whereas calcite, dolomite, and pyrite showed larger deviations. Phase errors reached up to 65.36% for dolomite, resulting in a weighted error of 26.74% and a spatial error of 49.20%. Using chemo-mechanical grid volume fractions (Fig. 12(b)) improved phase distribution and reduced the weighted error to 17.42%; however, the spatial error increased to 55.50% due to misclassification of minor phases. These results highlight the sensitivity of clustering to the reliability and representativeness of the input volume fractions, with non-representative SEM-EDS fractions leading to reduced clustering accuracy. To further improve spatial accuracy in future studies, automated grid alignment techniques or the use of reference markers could be explored to enhance the registration between mechanical testing locations and SEM-EDS maps. Despite these challenges, UMAP with k-means clustering remained relatively stable, producing converged and reproducible results across multiple random seeds with ARI values around 0.89. The technique maintained approximate characterization of dominant phases, even in more difficult settings like Test 3, where phase concentrations were lower and distribution less distinct.

Overall, the results suggest that UMAP with k-means can generalize reasonably well to independent grids, although performance remains dependent on the quality and representativeness of the chemo-mechanical reference data. The findings reaffirm that SEM-EDS image volume fractions are sufficient for approximate micromechanical characterization when sufficient correspondence exists between the imaging and indentation regions. Combined with results from Test 2, these findings support UMAP with k-means as a capable approach for high-resolution mechanical phase mapping in heterogeneous rocks using HSNM data.

Additional analysis on a reduced (smaller) indentation grid indicated that UMAP with k-means remained consistent even with fewer data points, achieving a spatial error of 31.00% and a weighted error of 20.83%. This suggests that clustering quality is more sensitive to the representativeness of the SEM-EDS image than the absolute size of the indentation grid.

# 5 CONCLUSIONS:

This study demonstrated that high-speed nanoindentation, combined with data-driven clustering and mineral phase volume fractions, can enable accurate micromechanical characterization and spatial mapping of heterogeneous rocks with reduced reliance on chemical analysis. Among the techniques evaluated, UMAP with k-means clustering consistently outperformed others, delivering the lowest weighted and spatial errors and producing modulus and hardness estimates that closely matched reference chemo-mechanical results for dominant phases.

The use of SEM-EDS image volume fractions—when representative of the indentation grid—was found to be sufficient for reliable clustering, providing a simplified alternative to chemo-mechanical preprocessing. Evaluation using an independent dataset confirmed the generalizability of the method, though accuracy diminished when mineral phases were sparsely distributed or poorly represented in the SEM-EDS image. Alternative techniques, such as DPMM, DBSCAN, and image segmentation provided approximate clustering performance and may serve as complementary validation tools, whereas GMM showed limited reliability under the tested conditions.

Overall, the findings support UMAP with k-means, paired with SEM-EDS image volume fractions, as a practical and scalable approach for microscale mechanical mapping in heterogeneous materials. While the present study focuses on Mancos shale, the UMAP with k-means framework is inherently data-driven and unsupervised, making it adaptable to other shales or heterogeneous geomaterials with different phase counts and microstructural complexity. Incorporating prior information, such as phase volume fractions, may further enhance clustering stability and phase proportion accuracy in more heterogeneous systems. Future work could explore validation across a broader range of geomaterials and investigate deep learning-based clustering methods to improve phase resolution in low-concentration regions, reduce manual intervention, and support ML-assisted image segmentation for more accurate results.

## 6 DATA AVAILABILITY STATEMENT:

Some or all data, models, or code that support the findings of this study are available from the corresponding author upon reasonable request.

## 7 ACKNOWLEDGEMENTS:

The authors would also like to acknowledge the financial support provided by the National Science Foundation (NSF) through Award 2045242, which enabled the completion of this project.

**TABLES:**

**Table 1.** Rock specimen composition along with their respective densities.

| Compound | Proportion (%) | Density (g/cm$^3$) |
|---|---|---|
| Clay (Illite) | 16.40 | 2.70 |
| Quartz | 33.20 | 2.65 |
| Feldspar | 4.70 | 2.56 |
| Calcite | 18.40 | 2.71 |
| Dolomite | 25.00 | 2.86 |
| Pyrite | 2.30 | 5.00 |

**Table 2.** Evaluation of optimal number of mineral phases

| Clusters | Test 2 | | | | Test 3 | | | |
|---|---|---|---|---|---|---|---|---|
| | SSE (x $10^5$) | Silhouette score | CHI | DBI | SSE (x $10^5$) | Silhouette score | CHI | DBI |
| 2 | 1.26 | 0.39 | 1260.42 | 1.11 | 1.06 | 0.42 | 1514.27 | 1.03 |
| 3 | 0.72 | 0.43 | 1831.34 | 0.80 | 0.63 | 0.44 | 1947.72 | 0.79 |
| 4 | 0.47 | 0.44 | 2255.09 | 0.74 | 0.44 | 0.42 | 2177.14 | 0.78 |
| 5 | 0.33 | 0.44 | 2621.89 | 0.72 | 0.35 | 0.39 | 2181.51 | 0.82 |
| 6 | 0.27 | 0.42 | 2651.98 | 0.76 | 0.29 | 0.38 | 2215.58 | 0.82 |

**Table 3.** Comparison of SEM-EDS image volume fractions and chemo-mechanical grid volume fractions for the Test 2 grid

| Cluster # | Mineral | Image Volume Fraction (in %) | Grid Volume Fraction (in %) | Error (%) |
|---|---|---|---|---|
| 1 | Clay (Illite) | 73.12 | 67.43 | 8.44 |
| 2 | Quartz | 16.67 | 23.08 | 27.77 |
| 3 | Feldspar | 1.39 | 1.73 | 19.65 |
| 4 | Calcite | 3.45 | 3.83 | 9.92 |
| 5 | Dolomite | 3.98 | 3.93 | 1.27 |
| 6 | Pyrite | 1.40 | 0.00 | 1.40 |
| Average | | | | 11.41 |

**Table 4.** Tabulated chemo-mechanical results for Test 2 grid

| Cluster # | Mean Modulus (GPa) | Mean Hardness (GPa) | No. of points | Expected Percentage of points |
|---|---|---|---|---|
| Clay | 27.94 | 1.39 | 1341 | 67.05 |
| Quartz | 67.77 | 8.82 | 454 | 22.70 |
| Feldspar | 61.56 | 6.97 | 27 | 1.35 |
| Calcite | 32.28 | 1.41 | 69 | 3.45 |
| Dolomite | 48.54 | 2.43 | 71 | 3.55 |
| Pyrite | 59.01 | 3.08 | 13 | 0.65 |
| Others | 25.23 | 0.79 | 25 | 1.25 |
| Total | | | 2000 | 100 |

**Table 5.** Test 2 grid UMAP with k-means clustering results with SEM-EDS image and chemo-mechanical grid volume fractions

| Type of Volume Fraction Used | Mineral Type | Mean Modulus (GPa) | Mean Hardness (GPa) | Computed Percentage of Points | Expected Percentage of points | Error (%) |
|---|---|---|---|---|---|---|
| Using SEM-EDS Image Volume fractions | Clay | 29.99 | 1.40 | 73.10 | 67.43 | 8.41 |
| | Quartz | 77.85 | 10.30 | 16.65 | 23.08 | 27.86 |
| | Feldspar | 109.75 | 17.41 | 2.50 | 1.73 | 44.51 |
| | Calcite | 11.85 | 0.25 | 3.30 | 3.83 | 13.84 |
| | Dolomite | 13.85 | 0.28 | 1.80 | 3.93 | 54.20 |
| | Pyrite | 8.12 | 0.14 | 2.65 | 0.00 | 2.65 |
| | Weighted Error | | | | | 13.40 |
| | Spatial Error | | | | | 30.40 |
| Using Chemo-Mechanical Grid Volume fractions | Clay | 24.15 | 0.92 | 67.40 | 67.43 | 0.04 |
| | Quartz | 57.37 | 5.22 | 23.05 | 23.08 | 0.13 |
| | Feldspar | 95.04 | 14.79 | 2.80 | 1.73 | 61.85 |
| | Calcite | 45.68 | 1.07 | 0.40 | 3.83 | 89.56 |
| | Dolomite | 82.69 | 12.78 | 2.95 | 3.93 | 24.94 |
| | Pyrite | 108.99 | 16.92 | 3.40 | 0.00 | 3.40 |
| | Weighted Error | | | | | 3.00 |
| | Spatial Error | | | | | 35.15 |

**Table 6.** Test 2 grid GMM clustering results with SEM-EDS image and chemo-mechanical grid volume fractions

| Type of Volume Fraction Used | Mineral Type | Mean Modulus (GPa) | Mean Hardness (GPa) | Computed Percentage of Points | Expected Percentage of points | Error (%) |
|---|---|---|---|---|---|---|
| Using SEM-EDS Image Volume fractions | Clay | 46.80 | 5.72 | 9.00 | 67.43 | 86.65 |
| | Quartz | 42.51 | 2.41 | 16.90 | 23.08 | 26.78 |
| | Feldspar | 90.59 | 14.18 | 10.80 | 1.73 | 524.28 |
| | Calcite | 18.92 | 0.42 | 32.50 | 3.83 | 748.56 |
| | Dolomite | 28.74 | 1.06 | 26.90 | 3.93 | 584.48 |
| | Pyrite | 87.89 | 8.13 | 3.90 | 0.00 | 3.90 |
| | Weighted Error | | | | | 469.61 |
| | Spatial Error | | | | | 81.30 |
| Using Chemo-Mechanical Grid Volume fractions | Clay | 29.49 | 1.13 | 26.80 | 67.43 | 60.26 |
| | Quartz | 87.78 | 7.07 | 2.80 | 23.08 | 87.87 |
| | Feldspar | 76.93 | 11.42 | 17.85 | 1.73 | 931.79 |
| | Calcite | 43.08 | 2.78 | 17.95 | 3.83 | 368.67 |
| | Dolomite | 19.25 | 0.44 | 34.60 | 3.93 | 780.41 |
| | Pyrite | - | - | 0.00 | 0.00 | 0.00 |
| | Weighted Error | | | | | 521.13 |
| | Spatial Error | | | | | 75.95 |

**Table 7.** Test 2 grid DPMM clustering results with SEM-EDS image and chemo-mechanical grid volume fractions

| Type of Volume Fraction Used | Mineral Type | Mean Modulus (GPa) | Mean Hardness (GPa) | Computed Percentage of Points | Expected Percentage of points | Error (%) |
|---|---|---|---|---|---|---|
| Using SEM-EDS Image Volume fractions | Clay | 21.58 | 0.60 | 50.35 | 67.43 | 25.33 |
| | Quartz | 88.42 | 13.63 | 12.25 | 23.08 | 46.92 |
| | Feldspar | 56.71 | 4.72 | 6.05 | 1.73 | 249.71 |
| | Calcite | 37.47 | 1.82 | 22.50 | 3.83 | 487.47 |
| | Dolomite | 40.30 | 5.13 | 6.35 | 3.93 | 61.58 |
| | Pyrite | 94.82 | 7.92 | 2.50 | 0.00 | 2.50 |
| | Weighted Error | | | | | 147.26 |
| | Spatial Error | | | | | 43.45 |
| Using Chemo-Mechanical Grid Volume fractions | Clay | 21.58 | 0.60 | 50.35 | 67.43 | 25.33 |
| | Quartz | 88.42 | 13.63 | 12.25 | 23.08 | 46.92 |
| | Feldspar | 56.71 | 4.72 | 6.05 | 1.73 | 249.71 |
| | Calcite | 37.47 | 1.82 | 22.50 | 3.83 | 487.47 |
| | Dolomite | 40.30 | 5.13 | 6.35 | 3.93 | 61.58 |
| | Pyrite | 94.82 | 7.92 | 2.50 | 0.00 | 2.50 |
| | Weighted Error | | | | | 147.26 |
| | Spatial Error | | | | | 43.45 |

**Table 8.** Test 2 grid DBSCAN clustering results with SEM-EDS image and chemo-mechanical grid volume fractions

| Type of Volume Fraction Used | Mineral Type | Mean Modulus (GPa) | Mean Hardness (GPa) | Computed Percentage of Points | Expected Percentage of points | Error (%) |
|---|---|---|---|---|---|---|
| Using SEM-EDS Image Volume fractions | Clay | 27.55 | 1.25 | 79.30 | 67.43 | 17.60 |
| | Quartz (noise included) | 84.90 | 11.44 | 17.65 | 23.08 | 23.53 |
| | Feldspar | 50.80 | 7.14 | 1.60 | 1.73 | 7.51 |
| | Calcite | 54.93 | 5.44 | 0.75 | 3.83 | 80.42 |
| | Dolomite | 61.37 | 4.36 | 0.70 | 3.93 | 82.19 |
| | Weighted Error | | | | | 19.41 |
| | Spatial Error | | | | | 31.70 |
| Using Chemo-Mechanical Grid Volume fractions | Clay | 27.55 | 1.25 | 79.30 | 67.43 | 17.60 |
| | Quartz (noise included) | 84.90 | 11.44 | 17.65 | 23.08 | 23.53 |
| | Feldspar | 50.80 | 7.14 | 1.60 | 1.73 | 7.51 |
| | Calcite | 54.93 | 5.44 | 0.75 | 3.83 | 80.42 |
| | Dolomite | 61.37 | 4.36 | 0.70 | 3.93 | 82.19 |
| | Weighted Error | | | | | 19.41 |
| | Spatial Error | | | | | 31.70 |

**Table 9.** Test 2 grid image processing and segmentation technique results with SEM-EDS image and chemo-mechanical grid volume fractions

| Type of Volume Fraction Used | Mineral Type | Mean Modulus (GPa) | Mean Hardness (GPa) | Computed Percentage of Points | Expected Percentage of points | Error (%) |
|---|---|---|---|---|---|---|
| Using SEM-EDS Image Volume fractions | Clay | 38.07 | 3.16 | 73.10 | 67.43 | 8.41 |
| | Quartz | 38.93 | 3.11 | 16.65 | 23.08 | 27.86 |
| | Feldspar | 46.49 | 4.09 | 1.40 | 1.73 | 19.08 |
| | Calcite | 35.72 | 2.94 | 3.45 | 3.83 | 9.92 |
| | Dolomite | 45.84 | 4.40 | 4.00 | 3.93 | 1.78 |
| | Pyrite | 32.74 | 1.99 | 1.40 | 0.00 | 1.40 |
| | Weighted Error | | | | | 11.49 |
| | Spatial Error | | | | | 46.60 |
| Using Chemo-Mechanical Grid Volume fractions | Clay | 37.80 | 3.11 | 67.45 | 67.43 | 0.03 |
| | Quartz | 39.86 | 3.35 | 23.05 | 23.08 | 0.13 |
| | Feldspar | 46.13 | 3.71 | 1.70 | 1.73 | 1.73 |
| | Calcite | 35.60 | 2.92 | 3.80 | 3.83 | 0.78 |
| | Dolomite | 41.92 | 3.80 | 3.95 | 3.93 | 0.51 |
| | Pyrite | 44.70 | 4.86 | 0.00 | 0.00 | 0.00 |
| | Weighted Error | | | | | 0.13 |
| | Spatial Error | | | | | 48.65 |

**Table 10.** Tabulated chemo-mechanical results for Test 3 grid

| Cluster # | Mean Modulus (GPa) | Mean Hardness (GPa) | No. of points | Expected Percentage of points |
|---|---|---|---|---|
| Clay | 28.66 | 1.79 | 1238 | 61.90 |
| Quartz | 36.69 | 2.93 | 184 | 9.20 |
| Feldspar | 35.80 | 2.87 | 21 | 1.05 |
| Calcite | 29.14 | 1.90 | 116 | 5.80 |
| Dolomite | 33.79 | 2.56 | 253 | 12.65 |
| Pyrite | 40.43 | 1.85 | 28 | 1.40 |
| Others | 28.93 | 1.94 | 160 | 8.00 |
| Total | | | 2000 | 100 |

**Table 11.** Comparison of SEM-EDS image volume fractions and chemo-mechanical grid volume fractions for the Test 3 grid

| Cluster # | Mineral | Image Volume Fraction (in %) | Grid Volume Fraction (in %) | Error (%) |
|---|---|---|---|---|
| 1 | Clay (Illite) | 76.10 | 63.25 | 20.32 |
| 2 | Quartz | 6.66 | 10.55 | 36.87 |
| 3 | Feldspar | 1.75 | 2.40 | 27.08 |
| 4 | Calcite | 3.99 | 7.10 | 43.80 |
| 5 | Dolomite | 8.92 | 14.00 | 36.29 |
| 6 | Pyrite | 2.60 | 2.70 | 3.70 |
| Average | | | | 28.01 |

**Table 12.** Test 3 grid UMAP with k-means clustering results with SEM-EDS image and chemo-mechanical grid volume fractions

| Type of Volume Fraction Used | Mineral Type | Mean Modulus (GPa) | Mean Hardness (GPa) | Computed Percentage of Points | Expected Percentage of points | Error (%) |
|---|---|---|---|---|---|---|
| Using SEM-EDS Image Volume fractions | Clay | 22.98 | 1.05 | 76.10 | 63.25 | 20.32 |
| | Quartz | 35.50 | 2.45 | 6.65 | 10.55 | 36.97 |
| | Feldspar | 34.93 | 0.83 | 3.50 | 2.40 | 45.83 |
| | Calcite | 49.55 | 1.90 | 4.65 | 7.10 | 34.51 |
| | Dolomite | 67.14 | 7.22 | 4.85 | 14.00 | 65.36 |
| | Pyrite | 87.07 | 14.03 | 4.25 | 2.70 | 57.41 |
| | Weighted Error | | | | | 26.74 |
| | Spatial Error | | | | | 49.20 |
| Using Chemo-Mechanical Grid Volume fractions | Clay | 20.32 | 0.72 | 63.25 | 63.25 | 0.00 |
| | Quartz | 37.62 | 2.70 | 10.55 | 10.55 | 0.00 |
| | Feldspar | 35.06 | 1.01 | 5.95 | 2.40 | 147.92 |
| | Calcite | 29.79 | 0.75 | 5.45 | 7.10 | 23.24 |
| | Dolomite | 57.43 | 5.20 | 9.75 | 14.00 | 30.36 |
| | Pyrite | 83.27 | 13.43 | 5.05 | 2.70 | 87.04 |
| | Weighted Error | | | | | 17.42 |
| | Spatial Error | | | | | 55.50 |

**FIGURE CAPTION LIST:**

Fig. 1. Proposed framework for micromechanical characterization of heterogeneous rocks

Fig. 2. Workflow for clustering-based micromechanical phase differentiation using HSNM data

Fig. 3. Indentation grid overlaid on a multispectral image showing the spatial distribution of mineral phases.

Fig. 4. Determination of the number of clusters for Test 2 and Test 3 grids using (a) SSE, (b) silhouette score, (c) CHI, and (d) DBI

Fig. 5. Chemo-mechanical results for the Test 2 grid: (a) 2D spatial map and (b) modulus versus hardness scatter plot

Fig. 6. Micromechanical characterization and mapping for Test 2 grid using UMAP with k-means clustering with (a) SEM-EDS image volume fractions: 2D spatial map (left) and modulus versus hardness scatter plot

(right) and (b) chemo-mechanical grid volume fractions: 2D spatial map (left) and modulus versus hardness scatter plot (right)

Fig. 7. Micromechanical characterization and mapping for Test 2 grid using GMM clustering with (a) SEM-EDS image volume fractions (random seed of 20): 2D spatial map (left) and modulus versus hardness scatter plot (right) and (b) chemo-mechanical grid volume fractions (random seed of 30): 2D spatial map (left) and modulus versus hardness scatter plot (right)

Fig. 8. Micromechanical characterization and mapping for Test 2 grid using DPMM clustering with (a) SEM-EDS image volume fractions (random seed of 60): 2D spatial map (left) and modulus versus hardness scatter plot (right) and (b) chemo-mechanical grid volume fractions (random seed of 60): 2D spatial map (left) and modulus versus hardness scatter plot (right)

Fig. 9. Micromechanical characterization and mapping for Test 2 grid using DBSCAN clustering with (a) SEM-EDS image volume fractions: 2D spatial map (left) and modulus versus hardness scatter plot (right) and (b) chemo-mechanical grid volume fractions: 2D spatial map (left) and modulus versus hardness scatter plot (right)

Fig. 10. Micromechanical characterization and mapping for Test 2 grid using image processing and segmentation technique with (a) SEM-EDS image volume fractions: 2D spatial map (left) and modulus versus hardness scatter plot (right) and (b) chemo-mechanical grid volume fractions: 2D spatial map (left) and modulus versus hardness scatter plot (right)

Fig. 11. Chemo-mechanical analysis of Test 3 grid: (a) 2D spatial map and (b) modulus versus hardness scatter plot

Fig. 12. Micromechanical characterization and mapping for Test 3 grid using UMAP with k-means clustering with (a) SEM-EDS image volume fractions: 2D spatial map (left) and modulus versus hardness scatter plot (right) and (b) chemo-mechanical grid volume fractions: 2D spatial map (left) and modulus versus hardness scatter plot (right)

**FIGURE FILES:**

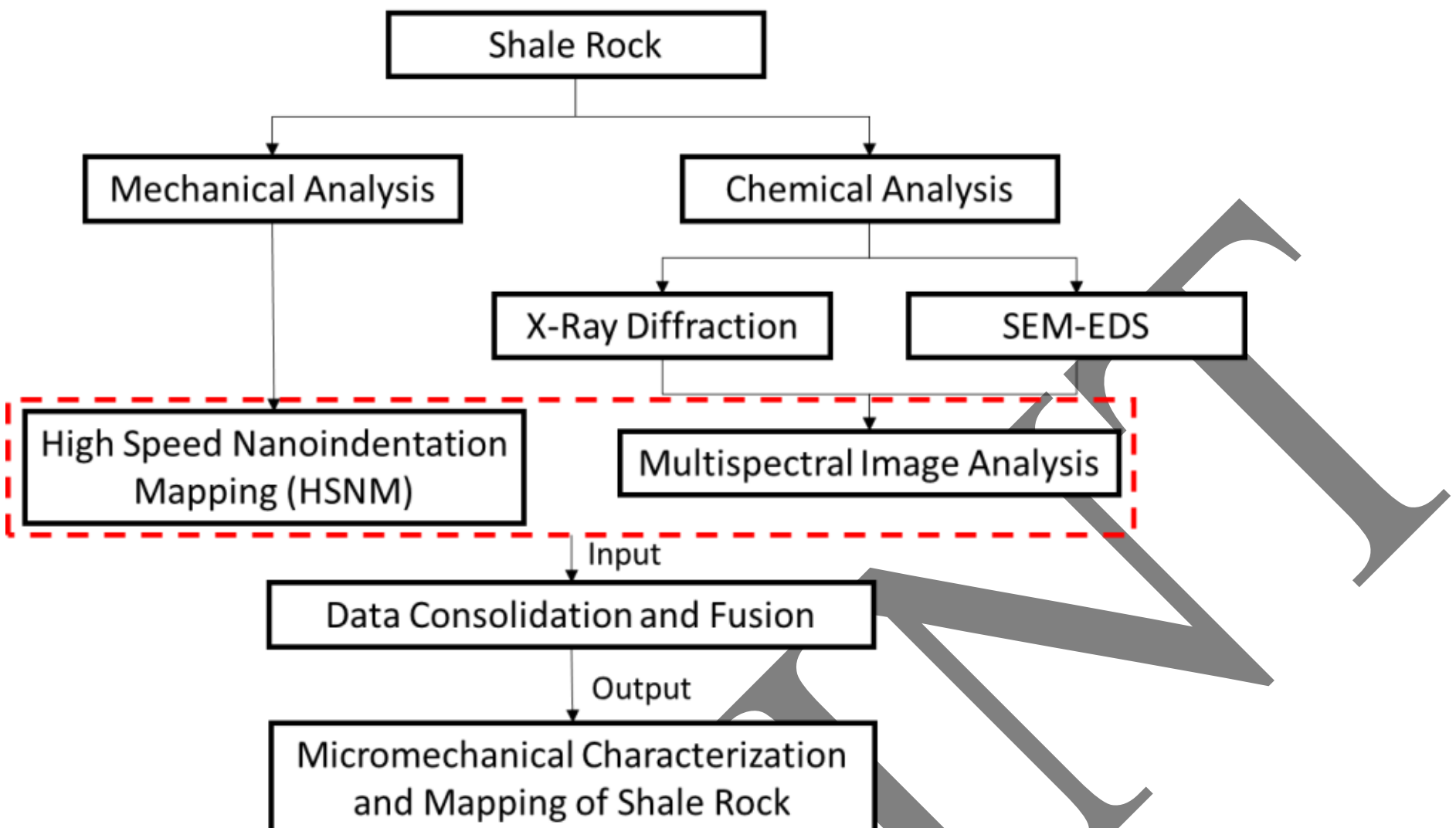

Fig. 1. Proposed framework for micromechanical characterization of heterogeneous rocks

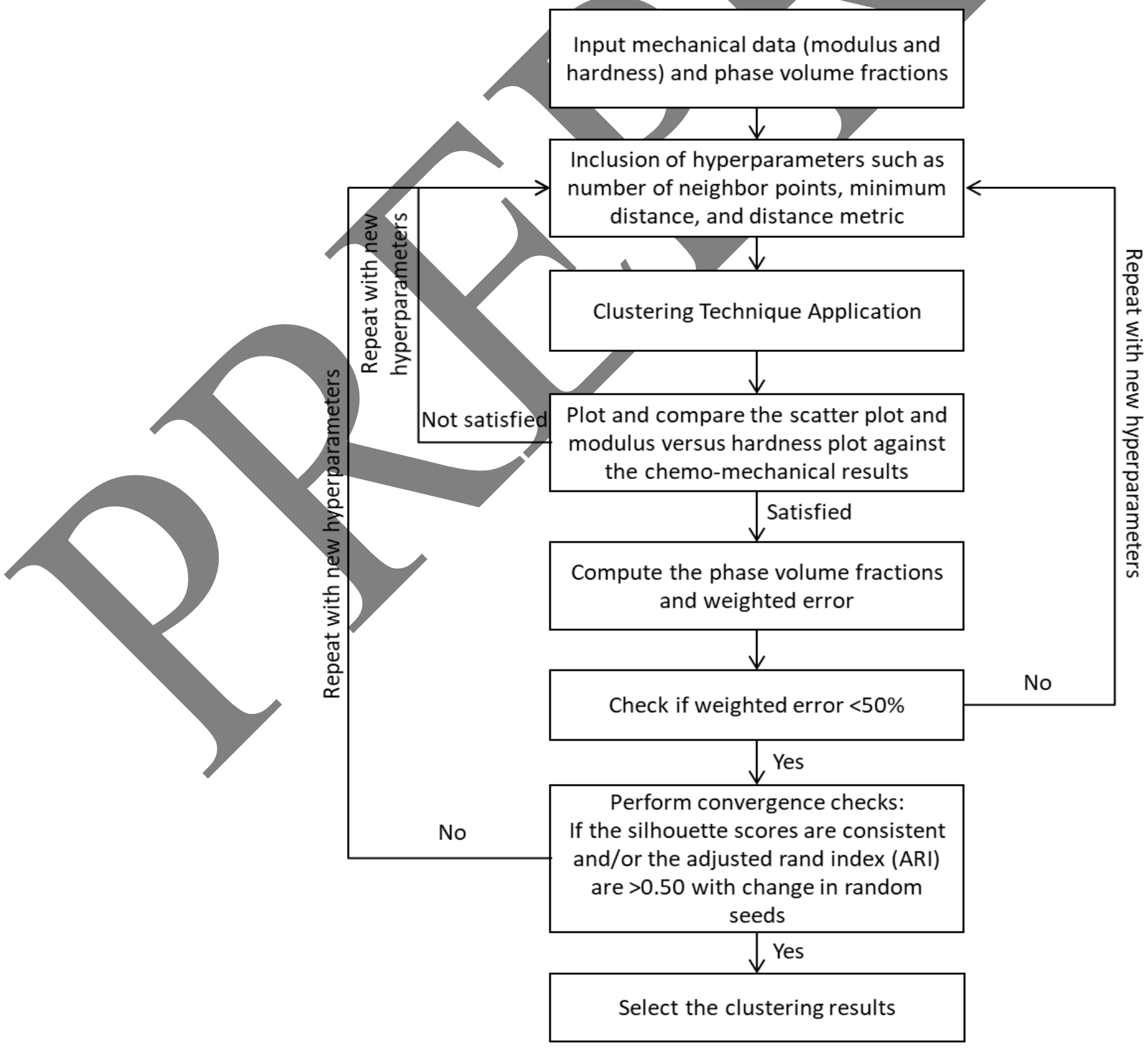

Fig. 2. Workflow for clustering-based micromechanical phase differentiation using HSNM data

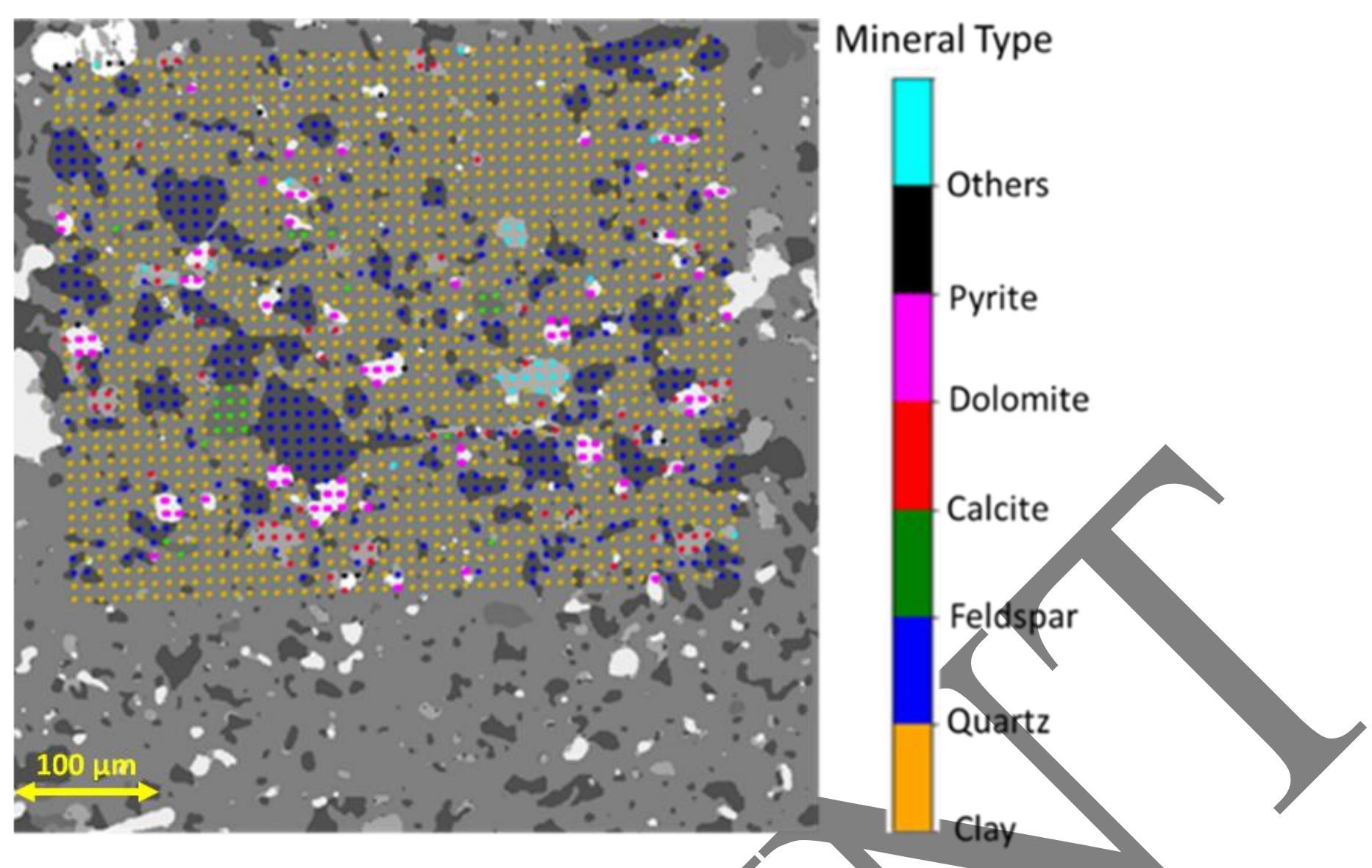

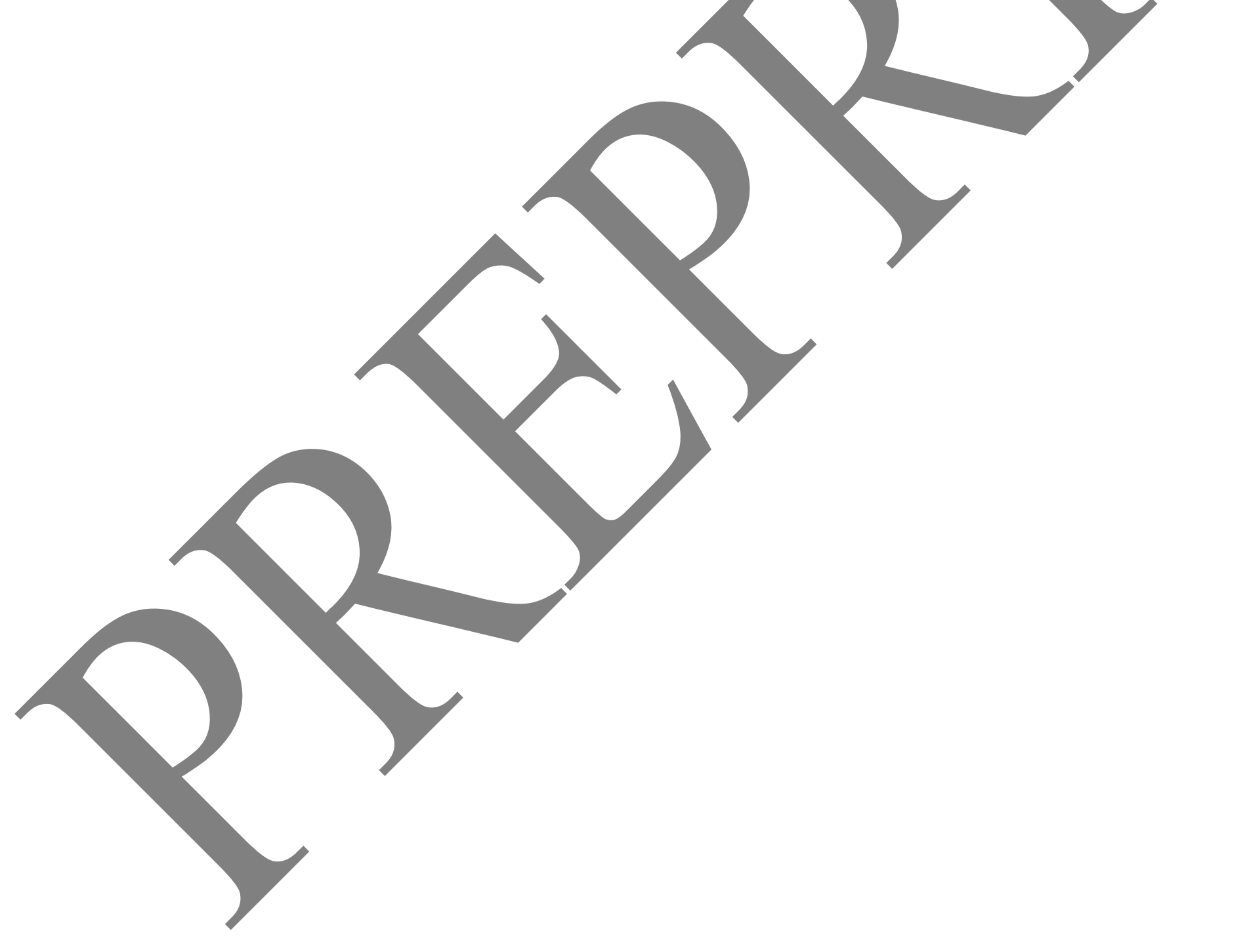

Fig. 3. Indentation grid overlaid on a multispectral image showing the spatial distribution of mineral phases.

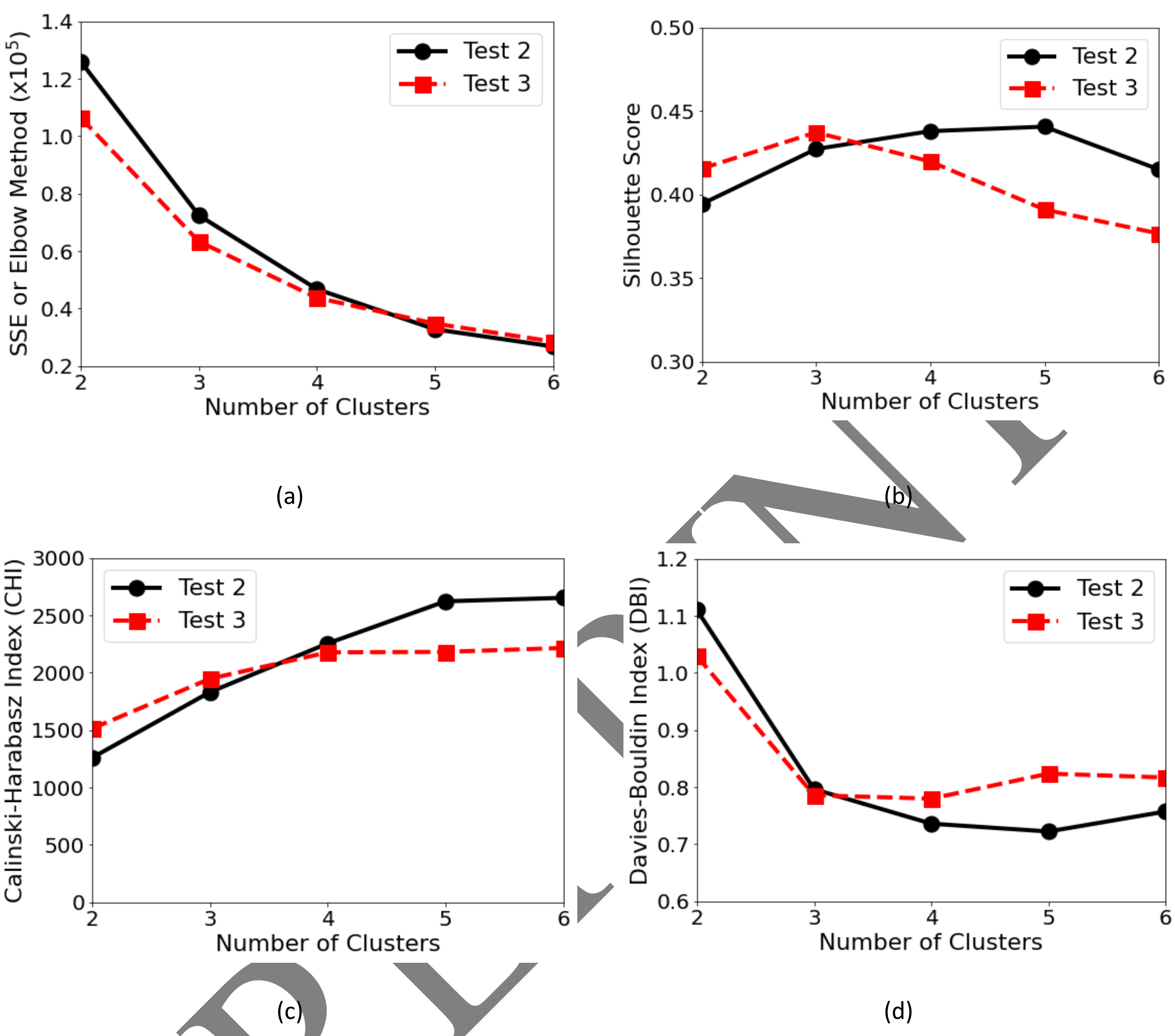

Fig. 4. Determination of the number of clusters for Test 2 and Test 3 grids using (a) SSE, (b) silhouette score, (c) CHI, and (d) DBI

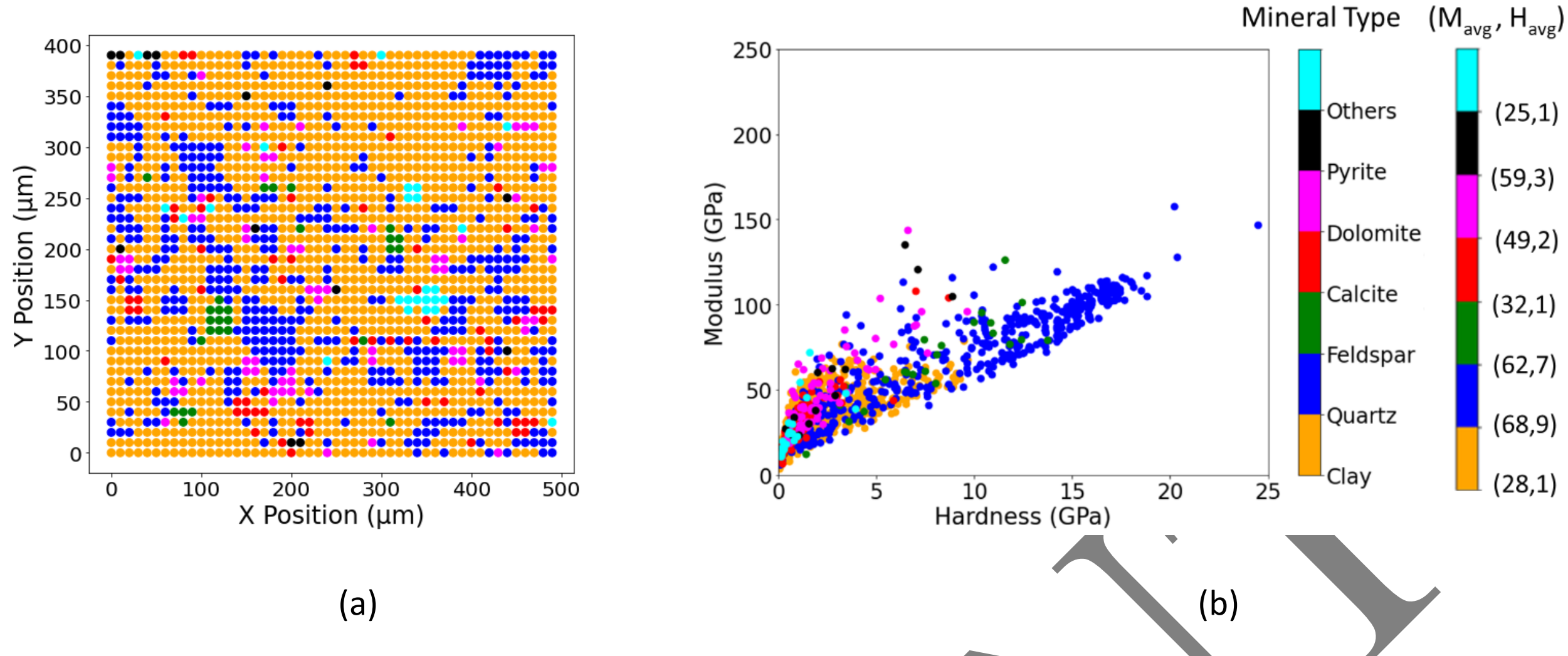

(a) (b)

Fig. 5. Chemo-mechanical results for the Test 2 grid: (a) 2D spatial map and (b) modulus versus hardness scatter plot

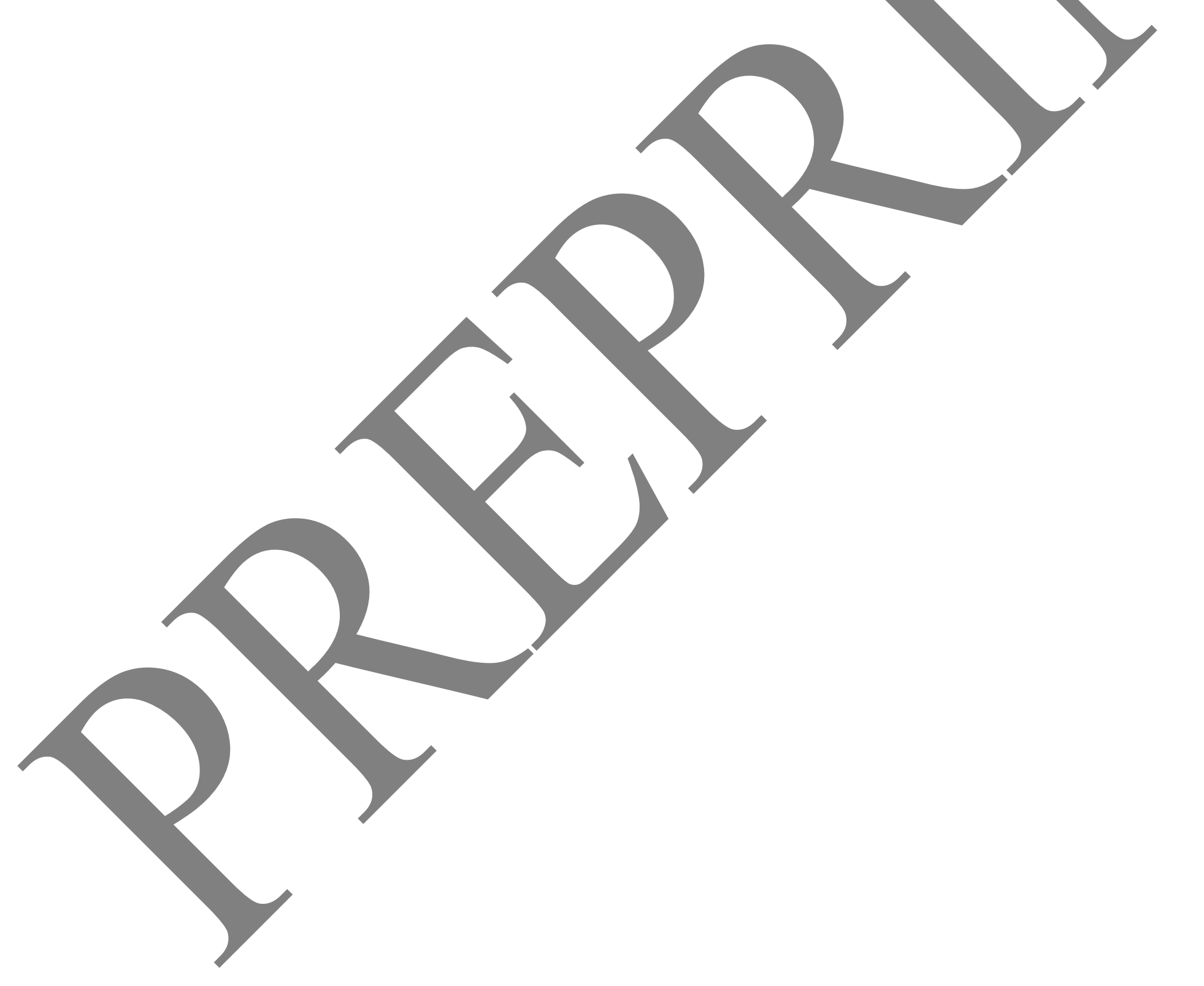

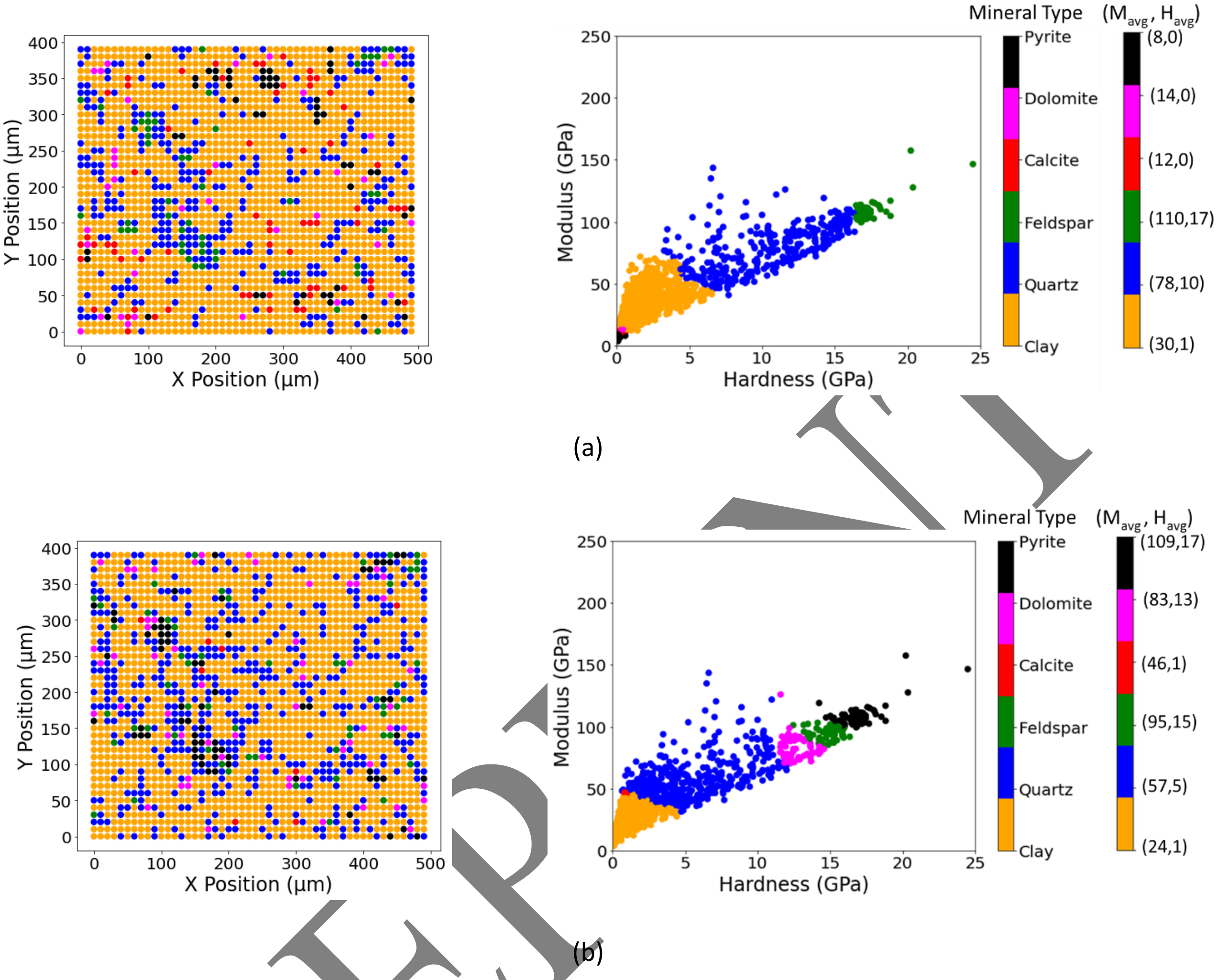

Fig. 6. Micromechanical characterization and mapping for Test 2 grid using UMAP with k-means clustering with (a) SEM-EDS image volume fractions: 2D spatial map (left) and modulus versus hardness scatter plot (right) and (b) chemo-mechanical grid volume fractions: 2D spatial map (left) and modulus versus hardness scatter plot (right)

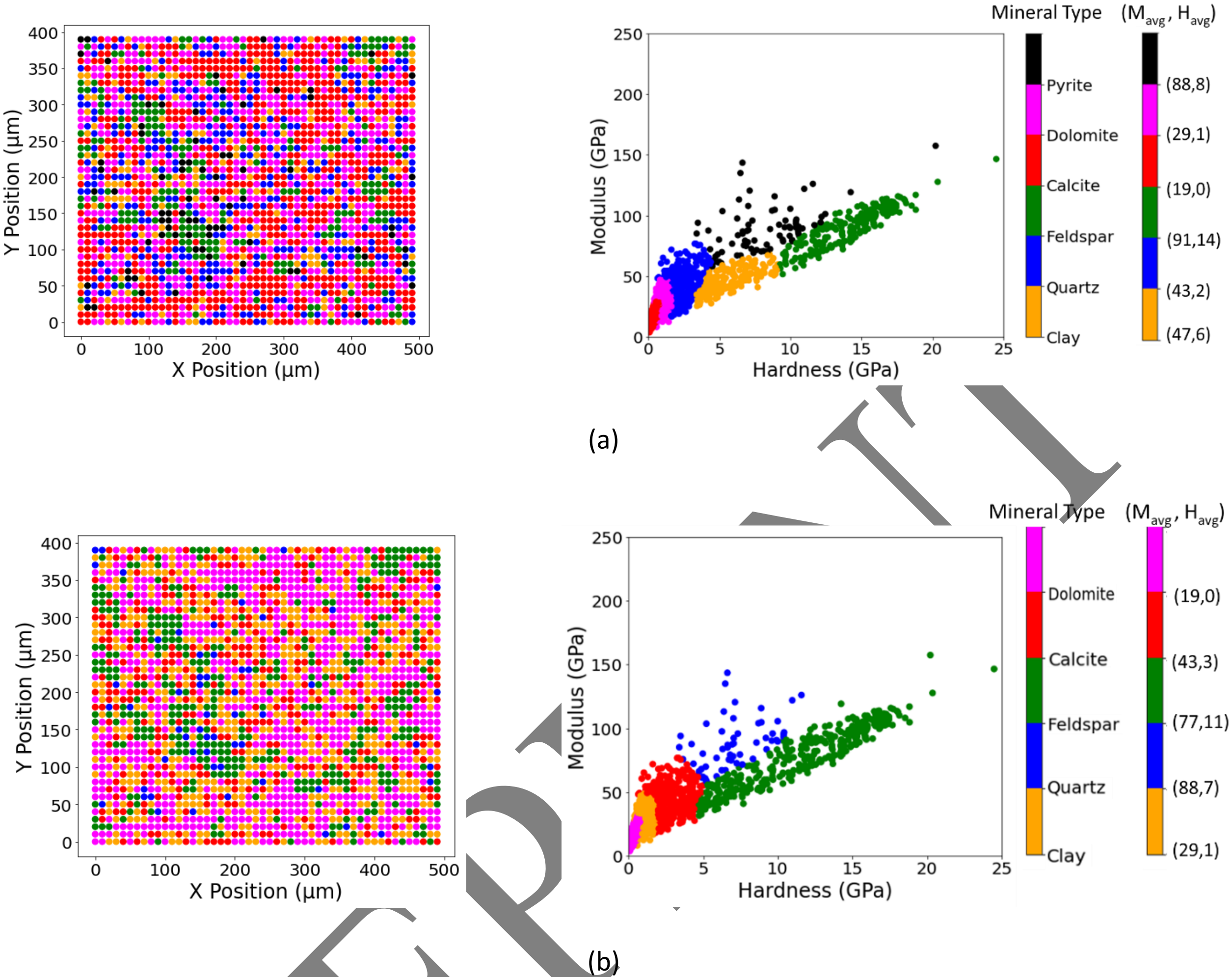

Fig. 7. Micromechanical characterization and mapping for Test 2 grid using GMM clustering with (a) SEM-EDS image volume fractions (random seed of 20): 2D spatial map (left) and modulus versus hardness scatter plot (right) and (b) chemo-mechanical grid volume fractions (random seed of 30): 2D spatial map (left) and modulus versus hardness scatter plot (right)

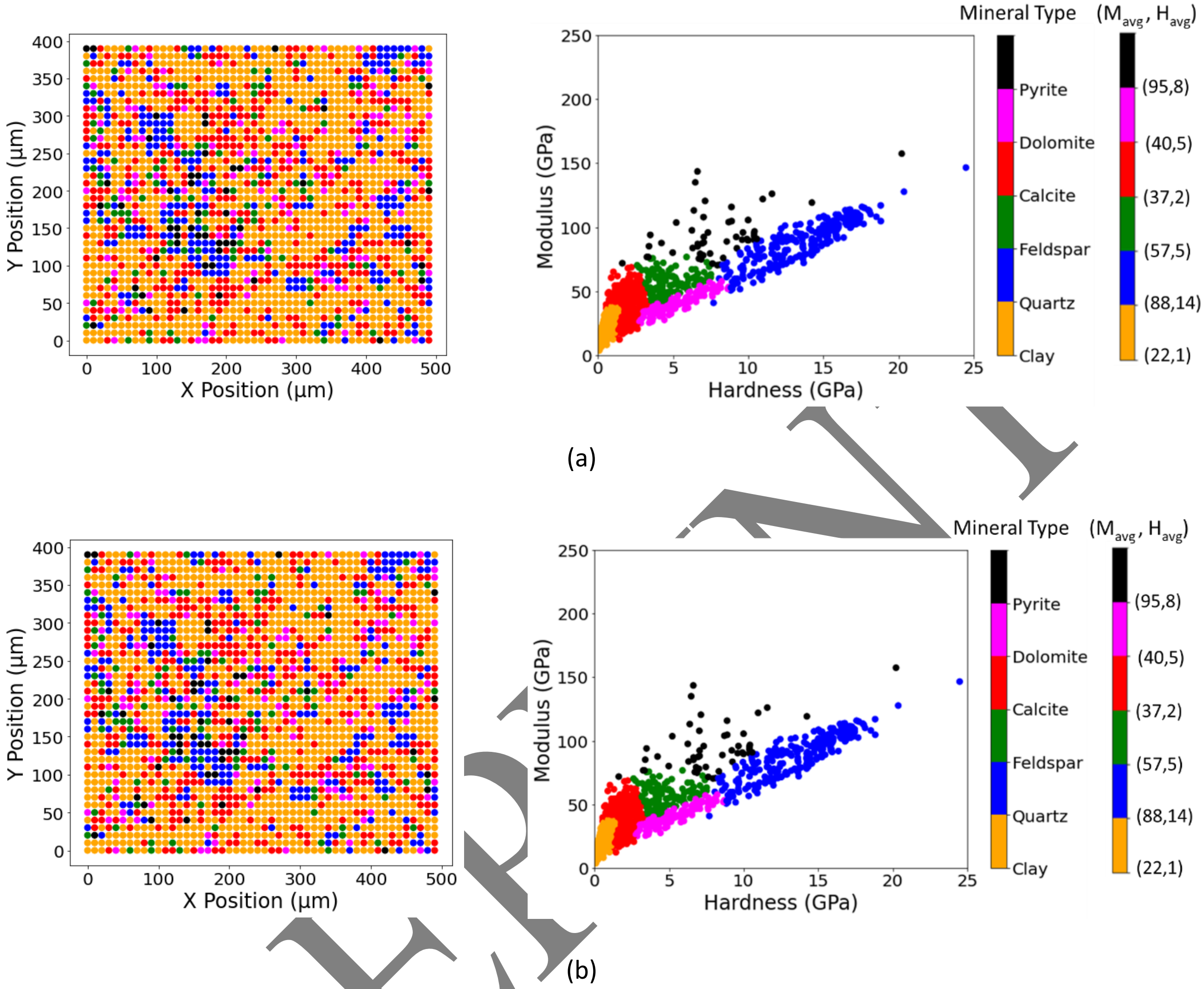

Fig. 8. Micromechanical characterization and mapping for Test 2 grid using DPMM clustering with (a) SEM-EDS image volume fractions (random seed of 60): 2D spatial map (left) and modulus versus hardness scatter plot (right) and (b) chemo-mechanical grid volume fractions (random seed of 60): 2D spatial map (left) and modulus versus hardness scatter plot (right)

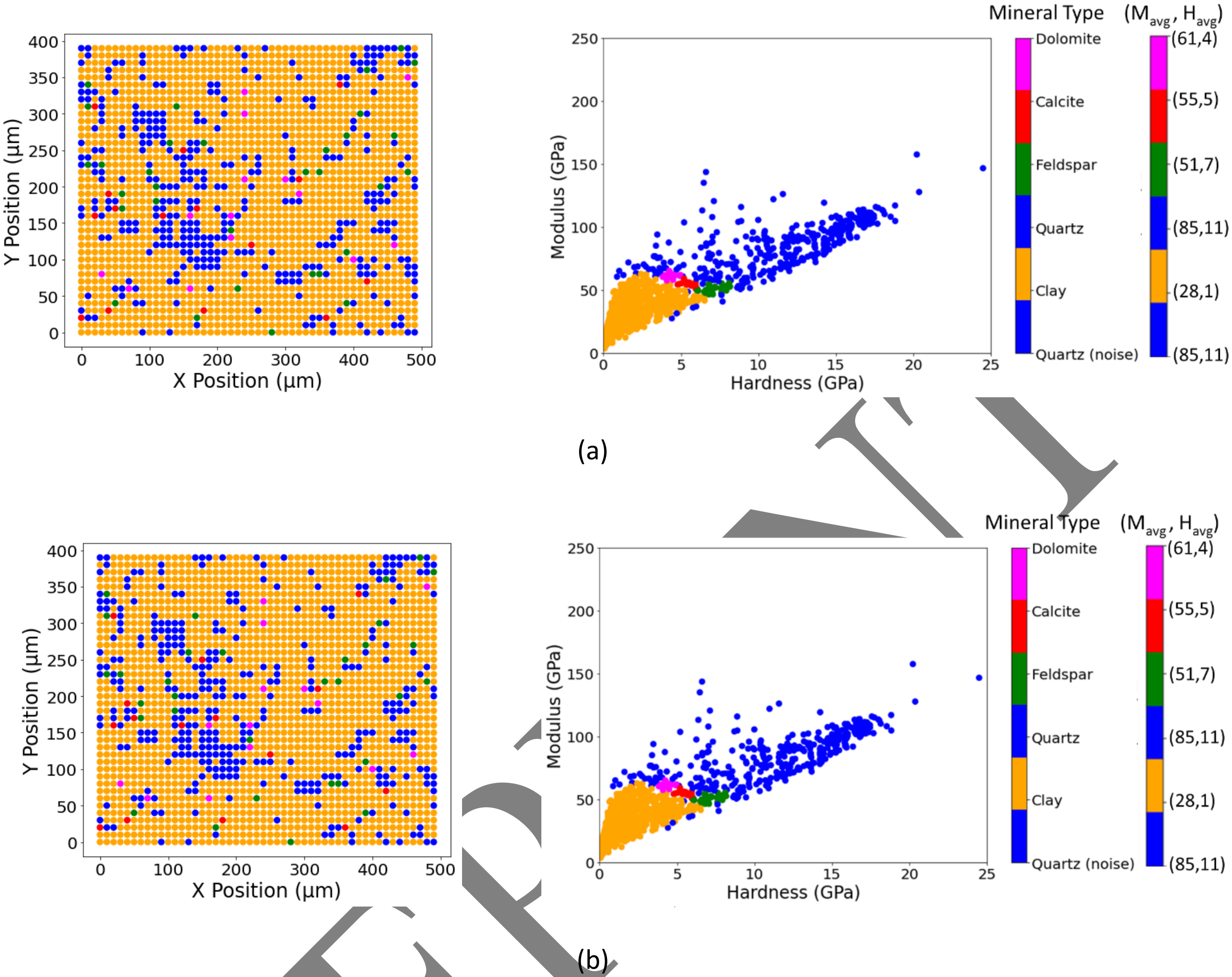

Fig. 9. Micromechanical characterization and mapping for Test 2 grid using DBSCAN clustering with (a) SEM-EDS image volume fractions: 2D spatial map (left) and modulus versus hardness scatter plot (right) and (b) chemo-mechanical grid volume fractions: 2D spatial map (left) and modulus versus hardness scatter plot (right)

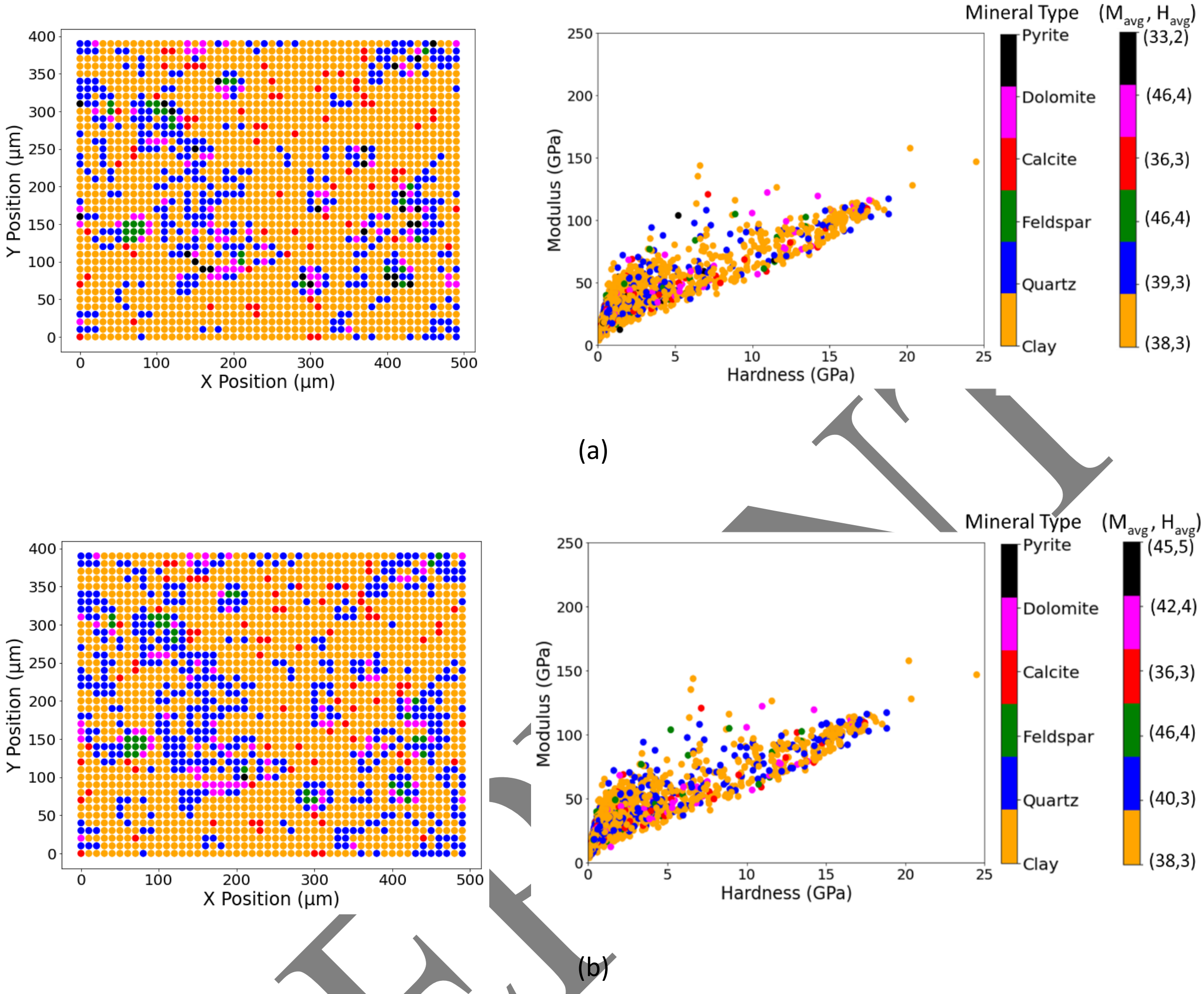

Fig. 10. Micromechanical characterization and mapping for Test 2 grid using image processing and segmentation technique with (a) SEM-EDS image volume fractions: 2D spatial map (left) and modulus versus hardness scatter plot (right) and (b) chemo-mechanical grid volume fractions: 2D spatial map (left) and modulus versus hardness scatter plot (right)

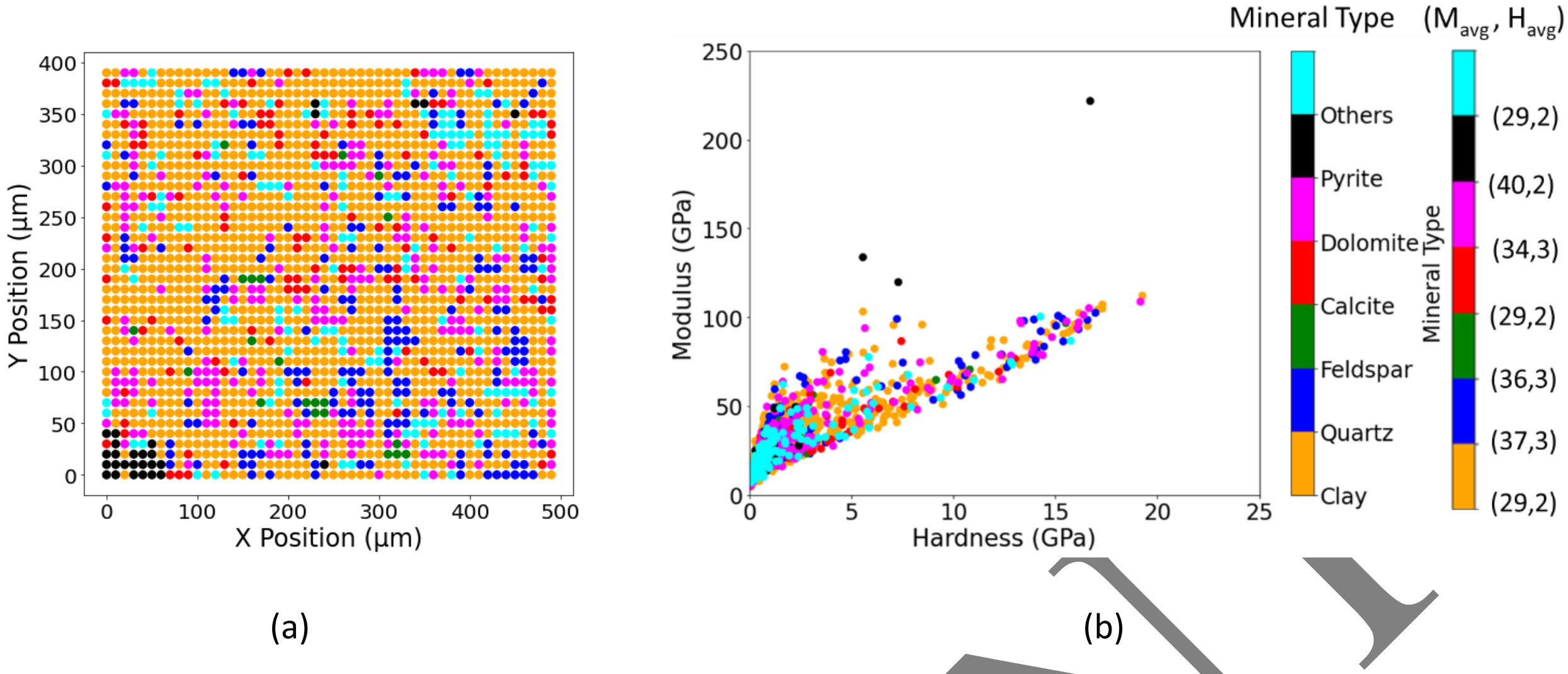

Fig. 11. Chemo-mechanical analysis of Test 3 grid: (a) 2D spatial map and (b) modulus versus hardness scatter plot

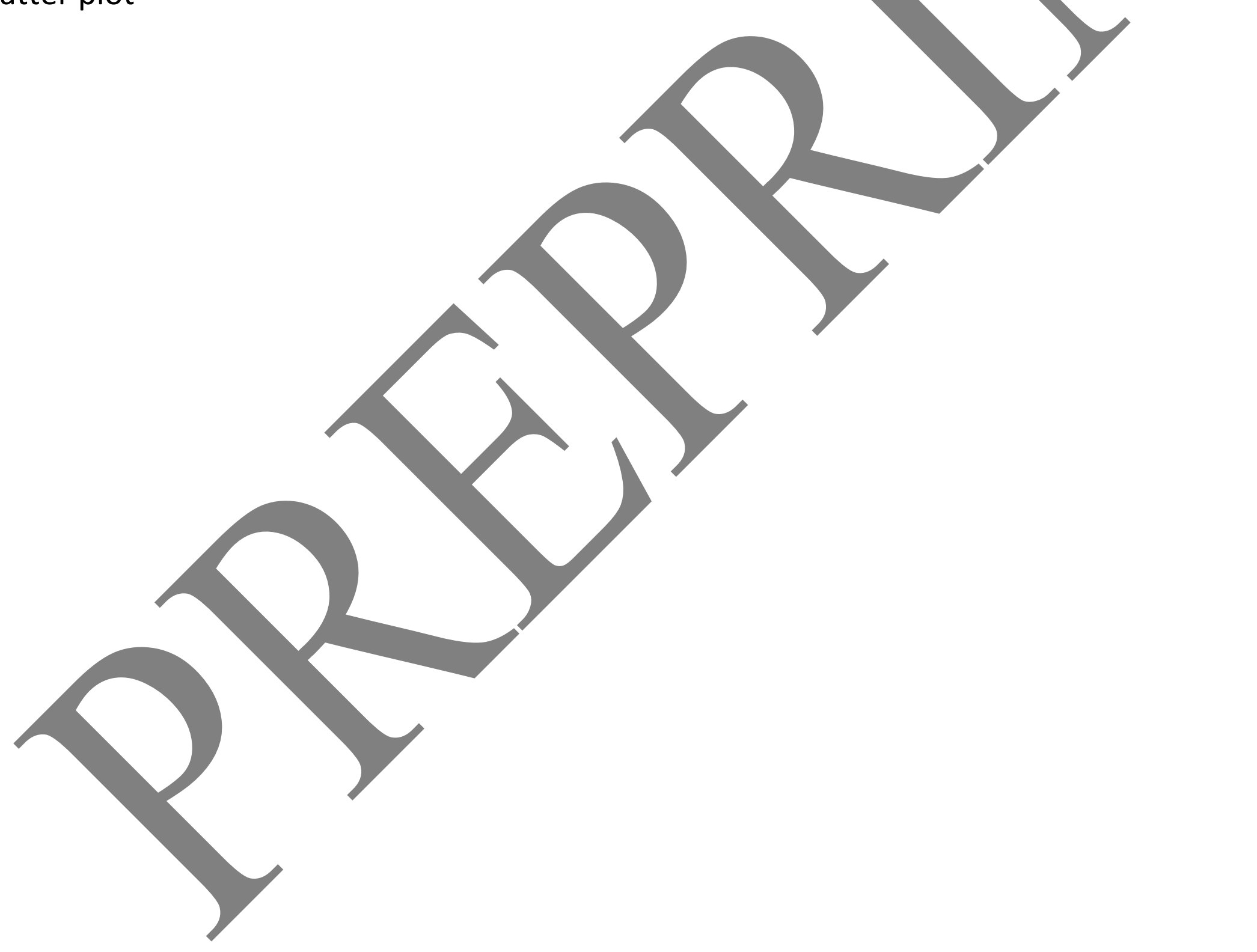

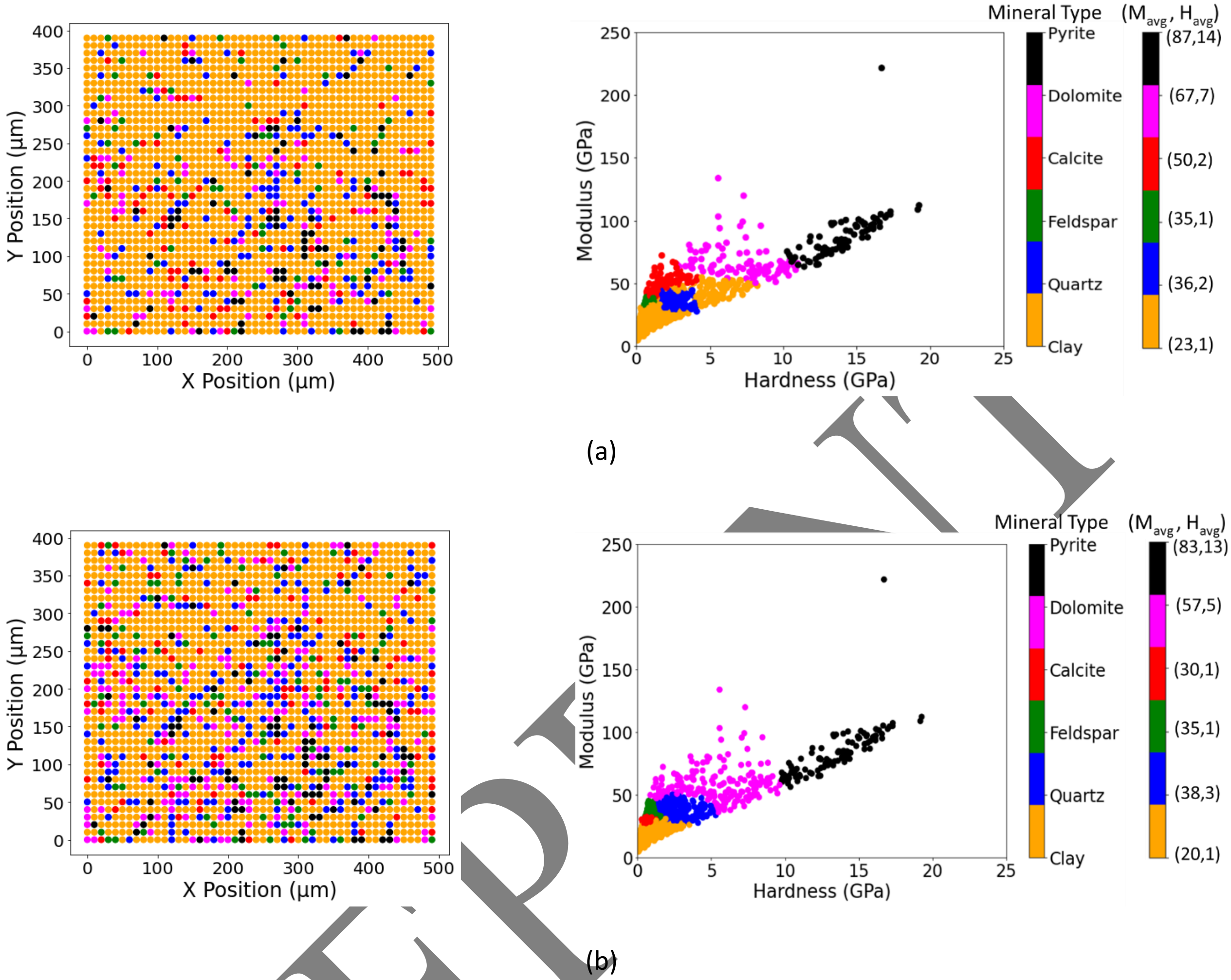

Fig. 12. Micromechanical characterization and mapping for Test 3 grid using UMAP with k-means clustering with (a) SEM-EDS image volume fractions: 2D spatial map (left) and modulus versus hardness scatter plot (right) and (b) chemo-mechanical grid volume fractions: 2D spatial map (left) and modulus versus hardness scatter plot (right)